\documentclass[aps,prd,nofootinbib,showkeys,preprint,floatfix]{revtex4-1}

\usepackage{graphicx}
\usepackage{graphics}
\usepackage[mathscr]{eucal}
\usepackage{amssymb}
\usepackage{amsmath}
\usepackage{xspace}
\usepackage{listings}
\usepackage{ulem}
\usepackage{color}

\newcommand{\sign}{\text{sign}\xspace}

\newcommand{\BL}{{\ensuremath{B-L}}\xspace}
\newcommand{\UBL}{{\ensuremath{U(1)_{B-L}}}\xspace}
\newcommand{\SUL}{{\ensuremath{SU(2)_{L}}}\xspace}
\newcommand{\BLSSM}{BLSSM\xspace}
\newcommand{\CBLSSM}{CBLSSM\xspace}
\newcommand{\msugra}{mSUGRA\xspace}

\newcommand{\EQ}{eq.\xspace}
\newcommand{\EQS}{eqs.\xspace}
\newcommand{\TAB}{Tab.\xspace}
\newcommand{\FIG}{Fig.\xspace}
\newcommand{\FIGS}{Figs.\xspace}
\newcommand{\EG}{\textit{e.g}.\xspace}
\newcommand{\IE}{\textit{i.e}.\xspace}
\newcommand{\REF}{Ref.\xspace}
\newcommand{\SEC}{sec.\xspace}
\def\gsim{\raise0.3ex\hbox{$\;>$\kern-0.75em\raise-1.1ex\hbox{$\sim\;$}}}

\newcommand{\gBL}[1]{{\ensuremath{g_{BL}^{#1}}}\xspace}
\newcommand{\gmix}{{\ensuremath{\bar{g}}}\xspace}

\newcommand{\rsnu}{R-sneutrino\xspace}
\newcommand{\rsnus}{R-sneutrinos\xspace}

\newcommand{\mRSq}{{\ensuremath{m_{{\tilde{\nu}}^{c}}^{2}}}\xspace}

\newcommand{\mZp}{{\ensuremath{m_{Z^{\prime}}}}\xspace}

\newcommand{\vev}[0]{VEV\xspace}
\newcommand{\vevs}[0]{VEVs\xspace}
\newcommand{\SPheno}[0]{\texttt{SPheno}\xspace}
\newcommand{\SARAH}[0]{\texttt{SARAH}\xspace}
\newcommand{\SSP}[0]{\texttt{SSP}\xspace}
\newcommand{\homps}[0]{\texttt{HOM4PS2}\xspace}
\newcommand{\RP}[0]{$R$-parity\xspace}
\newcommand{\RPVn}[0]{\RP violation\xspace}
\newcommand{\RPVg}[0]{\RP-violating\xspace}

\newcommand{\RPCg}[0]{\RP-conserving\xspace}

\newcommand{\AddrBonn}{%
Bethe Center for Theoretical Physics \& Physikalisches Institut der
 Universit\"at Bonn, \\
53115 Bonn, Germany }

\newcommand{\AddrWur}{%
Institut f\"ur Theoretische Physik und Astronomie,
Universit\"at W\"urzburg\\
Am Hubland,
97074 Wuerzburg}

\begin{document}

\title{Stability of $R$ parity in supersymmetric models extended By \UBL}

\author{J.\ E.\ Camargo-Molina} \email{jose.camargo@physik.uni-wuerzburg.de}

\author{B.\ O'Leary} \email{ben.oleary@physik.uni-wuerzburg.de}

\author{W.\ Porod} \email{porod@physik.uni-wuerzburg.de}\affiliation{\AddrWur}

\author{F.\ Staub}\email{fnstaub@th.physik.uni-bonn.de}
\affiliation{\AddrBonn}


\pacs{??, ??, ??}

\preprint{Bonn-TH-2012-033}
\begin{abstract}
We perform a study of the stability of \RPCg vacua of a constrained
 version of the minimal supersymmetric model with a gauged \UBL which can
 conserve \RP, using homotopy continuation to find all the extrema of the
 tree-level potential, for which we also calculated the one-loop corrections.
 While we find that a majority of the points in the parameter space preserve
 \RP, we find that a significant portion of points
 which na{\"{\i}}vely have phenomenologically acceptable vacua which
 conserve \RP actually have deeper vacua which break \RP through sneutrino
 \vevs. We investigate under what conditions the deeper \RPVg vacua appear.
 We find that while previous exploratory work was broadly correct in some of its
 qualitative conclusions, we disagree in detail.
\end{abstract}

\maketitle

\section{Introduction}

With the discovery of a resonance at 126 GeV at the LHC~\cite{:2012gk,:2012gu},
 the interpretation that it is the Higgs boson of the standard model (SM) raises
 the issue of the stability of the SM vacuum
\cite{Isidori:2001bm, Ellis:2009tp, EliasMiro:2011aa, Bezrukov:2012sa,
 Degrassi:2012ry}.
  Since there is only one field in the SM that can possibly have a non-zero
 vacuum expectation value (\vev) (assuming Lorentz invariance), finding the
 minima of the potential energy is straightforward, though of course evaluating
 it to the accuracy required is quite involved
\cite{EliasMiro:2011aa,Bezrukov:2012sa,Degrassi:2012ry}.

Many extensions of the SM introduce extra scalar fields, and thus could in
 principle have many more \vevs. In particular, the minimal supersymmetric
 extension of the SM (MSSM) not only introduces complex scalar partners for each
 SM fermion, it also introduces a second Higgs \SUL doublet. Even only
 allowing for the neutral components of the Higgs doublets to gain non-zero
 \vevs, the vacuum structure is non-trivial. Fortunately, as a particular case
 of a two-Higgs-doublet model, there is a proof that at tree level, there are at
 most two minima that conserve electric charge \cite{Ivanov:2006yq}, and
 there are analytic formulae to determine which vacuum has a lower energy at
 tree level \cite{Barroso:2012mj}.

However, the possibility of \vevs for the scalar partners of the SM fermions,
 the sfermions, is a troubling prospect, as all but the sneutrinos are charged
 under $SU(3)_{c}$ and/or $U(1)_{em}$, and thus any \vevs for these scalars
 are ruled out by these being good symmetries of the vacuum. Unfortunately, the
 potential as a function of several scalar fields is so complicated that even at
 tree level, finding the global minimum is highly non-trivial, in general
 \cite{Claudson:1983et,Drees:1985ie,Gunion:1987qv,Kuchimanchi:1993jg,Casas:1995pd}. The
 only tractable approaches thus far have been to investigate the ``most
 dangerous'' possibilities of stop or stau \vevs\
 \cite{Gunion:1987qv, Kitahara:2012pb, Carena:2012mw}.

Sneutrino \vevs, on the other hand, are not necessarily a 
problem\footnote{For example, in the 
case of bilinear \RPVn, the sneutrino
 \vevs have to be small to explain neutrino data implying that in this
 model essentially the conclusions concerning the minima of the potential
 can be taken over from the MSSM \cite{Hirsch:2004hr}.}. Of course,
 they would break \RP, leading to mixings between
SM and SUSY particles carrying the same spin and
 $SU(3)_c\times U(1)_{em}$ quantum numbers,
 see \EG~\cite{Hirsch:2000ef} and refs.\ therein.
However, in the absence of baryon-number-violating
 trilinear couplings in the superpotential, proton decay would still be
 perturbatively forbidden. If \RP is conserved, the lightest supersymmetric
 partner (LSP), if uncharged under $SU(3)_{c}$ and $U(1)_{em}$, could be a
 perturbatively stable dark matter candidate \cite{Ellis:1983ew,Ellis:2003cw}.
 If \RP is not conserved, the LSP
 is not absolutely stable, but, depending on the size of the sneutrino \vevs, a
 small amount of \RPVn still allows for a metastable LSP to make up the observed
 dark matter.
 However, even if the neutralino is
not sufficiently long-lived, another particle like the gravitino 
\cite{Borgani:1996ag,Takayama:2000uz,Hirsch:2005ag,Covi:2008jy,Steffen:2008qp}
 or the axion \cite{Rajagopal:1990yx,Goto:1991gq,Steffen:2008qp} could be
 long-lived enough to serve as dark matter.

The conservation of baryon number $B$ and of lepton number $L$ is only
 accidental in the SM, and this conservation through the imposition of \RP may
 appear to be rather \textit{ad hoc} in the MSSM, since the only difference
 between a lepton \SUL doublet and the Higgs \SUL doublet
 responsible for mass for the down-type quarks $H_{d}$ is that $H_{d}$ is
 assigned $L = 0$ instead of $1$. Still, noting that \RP is defined as
 $(-1)^{( 3 B + L - 2 s )} = (-1)^{( 3 ( B - L ) - 2 s )}$ if lepton number is
 integer or half-integer for every field, one may consider a symmetry based on
 $B-L$ as a motivated extension.

Adding \UBL as an extension to the SM
 \cite{Khalil:2006yi, Emam:2007dy,Basso:2008iv,Basso:2010yz,Basso:2010si} or the
 MSSM
 \cite{Khalil:2007dr,Barger:2008wn,FileviezPerez:2010ek}
 has been considered quite a lot in the literature, as in
 addition to motivating \RP, it leads to a rich phenomenology, and, in some
 models, naturally leads to type I see-saw mechanisms for neutrino masses.

The most minimal of such extensions of the MSSM requires superfields with
 right-handed neutrinos as fermionic components to cancel gauge anomalies, 
 and their scalar components provide the
 \vevs necessary to break \UBL as required by phenomenology. Already such a
 model has eight complex scalar fields that in principle could have non-zero
 \vevs without breaking $SU(3)_{c}$ or $U(1)_{em}$, with many possible minima.

Conserved \RP is still attractive from a phenomenological point of view, since
 there are tight limits on many \RPVg parameters
 \cite{Dreiner:2006gu,Dreiner:2010ye,Dreiner:2012mx}. One can
 extend the neutral sector with further superfields with \vevs which break \UBL
 without breaking \RP\ \cite{Khalil:2007dr, FileviezPerez:2010ek}. However,
 given the complicated structure of the
 potential, it is prudent to be concerned about whether such extensions really
 do conserve \RP at their global minima over relevant parameter ranges.

In the following sections, we investigate how stable \RPCg vacua are in such a
 minimal \UBL-gauged extension of the MSSM. In \SEC\ref{sec:model}, we define
 the model. In \SEC\ref{sec:methods}, we describe how we approach the difficult
 problem of minimizing such a complicated potential, including going to the full
 one-loop potential, since it is well-known that there can be rather large
 corrections in the Higgs sector of the MSSM
 \cite{Okada:1990vk,Okada:1990gg,Ellis:1990nz,Haber:1990aw,Ellis:1991zd,%
 Chankowski:1991md,Brignole:1992uf,Dabelstein:1994hb,Hempfling:1993qq,%
 Haber:1996fp,Heinemeyer:1998jw,Heinemeyer:1998kz,Heinemeyer:1998np,%
 Zhang:1998bm,Espinosa:1999zm,Espinosa:2001mm,Degrassi:2001yf,Brignole:2002bz,%
 Brignole:2001jy,Degrassi:2002fi,Martin:2002wn, Allanach:2004rh,%
 Harlander:2008ju,Kant:2010tf}.
 In \SEC~\ref{sec:results},
 we present the results of two scans over phenomenologically interesting regions
 of the parameter space, and in \SEC~\ref{sec:fileviezspinnercomparison}, we
 compare our results to previous investigations in the literature. Finally, we
 sum up in \SEC~\ref{sec:conclusion}.

\section{The model}
\label{sec:model}

\subsection{Particle content and superpotential}

There are several ways to extend the MSSM by \UBL. Here we choose the minimal
 extension which allows for a spontaneously broken \UBL without necessarily
 breaking \RP. This requires the addition of two SM gauge-singlet chiral
 superfields carrying \BL which may develop \vevs, as well as the addition of
 three generations of superfields containing right-handed neutrinos. We refer to
 this model as the \BLSSM.

The \BLSSM appears to be a relatively straightforward extension of the
 MSSM, yet it has a rich phenomenology, with a $Z'$ boson (with prospects for
 the LHC discussed in \cite{Krauss:2012ku}), Majorana neutrinos with see-saw
 masses, several qualitatively new dark matter candidates with respect to the
 MSSM \cite{Basso:2012gz}, and a rich Higgs boson sector
 \cite{Basso:2012tr}. It also has many interesting
 technicalities~\cite{Fonseca:2011vn}, which are fully explained in
 \cite{OLeary:2011yq}.

However, all the above works assume that \RP is conserved. Indeed, the model was
 constructed as the minimal extension of the MSSM that includes \UBL as an extra
 gauge symmetry, while allowing it to break spontaneously without necessarily
 violating the conservation of \RP. As mentioned in the introduction, \vevs
 for the scalar partners of the right-handed neutrinos could break \UBL at the
 cost of losing \RP, and aspects of the LHC phenomenology in the case of
 broken \RP are discussed in \cite{FileviezPerez:2011kd,FileviezPerez:2012mj}.
 The purpose of this work is to investigate how
 robustly \RP is conserved for parameters of phenomenological interest.

While we refer the reader to \REF~\cite{OLeary:2011yq} for a full
 description of the \BLSSM and its implementation in
 \SARAH\ \cite{Staub:2008uz,Staub:2009bi,Staub:2010jh,Staub:2012pb}
 and \SPheno\ \cite{Porod:2003um,Porod:2011nf}, for
 reference we recap the gauge symmetries and particle content of the model here.

The model consists of three generations of matter particles including
right-handed neutrinos which can, for example, be embedded in $SO(10)$
16-plets. For convenience, we refer to their scalar components as \rsnus.
 Moreover, below the GUT scale the usual MSSM Higgs doublets
are present, as well as two fields $\eta$ and $\bar{\eta}$ responsible
for the breaking of the \UBL. Furthermore, $\eta$ is responsible for
generating a Majorana mass term for the right-handed neutrinos and
thus we interpret the \BL charge of this field as its lepton number,
and likewise for $\bar{\eta}$, and call these fields bileptons since
they carry twice the lepton number of (anti-)neutrinos.  We summarize
the quantum numbers of the chiral superfields with respect to the
 model's gauge group $U(1)_Y \times SU(2)_L \times SU(3)_C \times \UBL$ in
 \TAB~\ref{tab:cSF}.
\begin{table} 
\centering
\begin{tabular}{|c|c|c|c|c|c|} 
\hline \hline 
Superfield & Spin 0 & Spin \(\frac{1}{2}\) & Generations & \, \( U(1)_Y\otimes\,
SU(2)_L\otimes\, SU(3)_C\otimes\, \UBL \) \, \\ 
\hline 
\(\hat{Q}\) & \(\tilde{Q}\) & \(Q\) & 3
 & \((\frac{1}{6},{\bf 2},{\bf 3},\frac{1}{6}) \) \\ 
\(\hat{d}^c\) & \(\tilde{d}^c\) & \(d^c\) & 3
 & \((\frac{1}{3},{\bf 1},{\bf \overline{3}},-\frac{1}{6}) \) \\ 
\(\hat{u}^c\) & \(\tilde{u}^c\) & \(u^c\) & 3
 & \((-\frac{2}{3},{\bf 1},{\bf \overline{3}},-\frac{1}{6}) \) \\ 
\(\hat{L}\) & \(\tilde{L}\) & \(L\) & 3
 & \((-\frac{1}{2},{\bf 2},{\bf 1},-\frac{1}{2}) \) \\ 
\(\hat{e}^c\) & \(\tilde{e}^c\) & \(e^c\) & 3
 & \((1,{\bf 1},{\bf 1},\frac{1}{2}) \) \\ 
\(\hat{\nu}^c\) & \(\tilde{\nu}^c\) & \(\nu^c\) & 3
 & \((0,{\bf 1},{\bf 1},\frac{1}{2}) \) \\ 
\(\hat{H}_d\) & \(H_d\) & \(\tilde{H}_d\) & 1
 & \((-\frac{1}{2},{\bf 2},{\bf 1},0) \) \\ 
\(\hat{H}_u\) & \(H_u\) & \(\tilde{H}_u\) & 1
 & \((\frac{1}{2},{\bf 2},{\bf 1},0) \) \\ 
\(\hat{\eta}\) & \(\eta\) & \(\tilde{\eta}\) & 1
 & \((0,{\bf 1},{\bf 1},-1) \) \\ 
\(\hat{\bar{\eta}}\) & \(\bar{\eta}\) & \(\tilde{\bar{\eta}}\) & 1
 & \((0,{\bf 1},{\bf 1},1) \) \\ 
\hline \hline
\end{tabular} 
\caption{Chiral superfields and their quantum numbers.}
\label{tab:cSF}
\end{table}

The superpotential is given by 
\begin{align} 
\nonumber 
W = & \, Y^{ij}_u\,\hat{u}^c_i\,\hat{Q}_j\,\hat{H}_u\,
- Y_d^{ij} \,\hat{d}^c_i\,\hat{Q}_j\,\hat{H}_d\,
- Y^{ij}_e \,\hat{e}^c_i\,\hat{L}_j\,\hat{H}_d\,+\mu\,\hat{H}_u\,\hat{H}_d\, \\
 & \, \, 
+Y^{ij}_{\nu}\,\hat{\nu}^c_i\,\hat{L}_j\,\hat{H}_u\,
- \mu' \,\hat{\eta}\,\hat{\bar{\eta}}\,
+Y^{ij}_x\,\hat{\nu}^c_i\,\hat{\eta}\,\hat{\nu}^c_j\,
\label{eq:superpot},
\end{align} 
and we have the additional soft SUSY-breaking terms:
\begin{align}
\nonumber \mathscr{L}_{SB} = & \mathscr{L}_{MSSM}
 - m_{\eta}^2 |\eta|^2 - m_{\bar{\eta}}^2 |\bar{\eta}|^2
 - {m_{\nu^c,ij}^{2}} (\tilde{\nu}_i^c)^* \tilde{\nu}_j^c \\
&+ \left( T^{ij}_{\nu}  H_u \tilde{\nu}_i^c \tilde{L}_j
 + T^{ij}_{x} \eta \tilde{\nu}_i^c \tilde{\nu}_j^c
 - \lambda_{\tilde{B}} \lambda_{\tilde{B}'} {M}_{B B'}
 - \frac{1}{2} \lambda_{\tilde{B}'} \lambda_{\tilde{B}'} {M}_{B'}
 - \eta \bar{\eta} B_{\mu'} + c.c\right) ,
\end{align}
with $i,j$ being the generation indices. Without loss of generality one can take
$B_\mu$ and $B_{\mu'}$ to be real. The extended gauge group breaks to
$SU(3)_C \otimes U(1)_{em}$ as the Higgs fields and bileptons receive
vacuum expectation values (\vevs):
\begin{align} 
H_d^0 = & \, \frac{1}{\sqrt{2}} \left(v_d  + \sigma_{d} + i \phi_{d} \right),
\hspace{1cm}
H_u^0 = \, \frac{1}{\sqrt{2}} \left(v_u  + \sigma_{u} + i \phi_{u} \right),\\ 
\eta
= & \, \frac{1}{\sqrt{2}} \left(v_{\eta} + \sigma_\eta + i \phi_{\eta} \right),
\hspace{1cm}
\bar{\eta}
= \, \frac{1}{\sqrt{2}} \left(v_{\bar{\eta}} + \sigma_{\bar{\eta}}
 + i \phi_{\bar{\eta}} \right).
\end{align}
We define $\tan\beta' = v_{\eta}/v_{\bar{\eta}}$ in analogy to
the ratio of the MSSM \vevs ($\tan\beta = v_{u}/v_{d}$).
For certain parameter combinations a spontaneous breakdown of
 \RP can occur as also
 some or all sneutrinos can obtain non-vanishing \vevs \cite{FileviezPerez:2010ek}.
We denote the \vevs for the sneutrinos of the
 \SUL doublets $\tilde L_{i}$ by $v_{L,i}$ and those of the \SUL
 singlet sneutrinos $\tilde \nu^c_{i}$ by $v_{R,i}$, with $i=1, 2, 3$.

\subsection{Constrained model}

We will consider in the following a scenario motivated by minimal
supergravity assuming a GUT unification of all soft SUSY-breaking scalar mass
 parameters as well as a unification of all gaugino mass parameters:
\begin{align}
 m^2_0 = & m^2_{H_d} = m^2_{H_u} = m^2_{\eta} = m^2_{\bar{\eta}}, \\
m^2_0 \mathbf{1} = & m_D^2  =  m_U^2  = m_Q^2 
= m_E^2  = m_L^2  = \mRSq , \\
 M_{1/2} = & M_1 = M_2 = M_3 = M_{\tilde{B}'}.
\end{align}
Also, for the trilinear soft SUSY-breaking couplings, we assume the ordinary
 \msugra-inspired conditions
\begin{align}
 T_{i} = A_{0} Y_{i}, \hspace{1cm} i = e,d,u,x,\nu \thickspace . 
\end{align}
Furthermore, we assume that there are no off-diagonal gauge couplings
or off-diagonal gaugino mass parameters present at the GUT scale, motivated by
 the possibility that the two Abelian groups are a
 remnant of a larger product group which gets broken at the GUT scale,
 as discussed in \cite{OLeary:2011yq}, to which we refer the reader for details
 of the gauge coupling structure.

In addition, we consider the mass of the $Z'$ and $\tan\beta'$ as 
 inputs and use the following set of free parameters:
\begin{eqnarray}
& m_0, \thickspace M_{1/2},\thickspace A_0,\thickspace \tan\beta,\thickspace
 \tan\beta',\thickspace \sign(\mu),\thickspace \sign(\mu'),\thickspace \mZp,
 \thickspace  Y_x \thickspace \mbox{and} \thickspace Y_{\nu} . &
\end{eqnarray}
\(Y_{\nu}\) is constrained by neutrino data and must therefore be
 very small in comparison to the other couplings in this model, 
 as required by the embedded TeV-scale type-I seesaw mechanism, so we take
 $Y_{\nu}^{ij} = 10^{-5} {\delta}^{ij}$
  as its precise structure does not affect any of our conclusions.
   $Y_x$ can always
 be taken diagonal and thus effectively we have 9 free parameters and 2 signs.
 We denote this constrained version of the \BLSSM as the \CBLSSM.

\section{Minimizing the potential}
\label{sec:methods}

The primary question that we wish to answer is: within the \CBLSSM, what portion
 of the parameter points which have a phenomenologically acceptable, \RPCg local
 minimum of the potential turn out to have a global minimum with different
 \vevs, and are there any patterns in the parameters of such points? By
 ``phenomenologically acceptable'', we restrict ourselves to parameter points
 which have a local minimum with the expectation values
 of the Higgs doublets leading to the correct
 values for $m_W$ and $m_Z$, there are no
 tachyons, and no charged or colored scalar has a \vev. Hence any minima that
 are ``phenomenologically acceptable'' and \RPCg have expectation
 values for the Higgs doublet fields $H_{d}, H_{u}$ and the bilepton fields
 $\eta, \bar{\eta}$, and no other field has a non-zero expectation
 value.

Our method to address this question was as follows: we performed a random scan
 over a range of input parameters, constrained to having phenomenologically
 acceptable, \RPCg local minima at expectation values for the Higgs
 doublet and bilepton fields given as input; we then found all the extrema of
 the tree-level potential for each parameter point allowing non-zero sneutrino
 \vevs;
 we evaluated the loop-corrected potential at these points (ensuring that the
 corrections did not move the minima significantly) and selected the deepest,
 and thus can say if the minimum chosen as input was not the global minimum.

\subsection{Generation of parameter points}

The package \SARAH was used to create a \SPheno executable specific to the
 \CBLSSM, which was used with the \SSP package \cite{Staub:2011dp} to perform
 random scans over two different parameter regions. 
The first scan, which we refer to as the
 ``democratic'' scan, took random values for each diagonal entry of the
 \rsnu\ -- bilepton Yukawa coupling $Y_{x}$ independently over its range. The
 other,
 which we refer to as the ``hierarchical'' scan, kept the $( 1, 1 )$ and
 $( 2, 2 )$ entries as $10^{-3}$ and $10^{-2}$ respectively. The hierarchical
 scan was motivated by the hierarchy of the quark and charged lepton Yukawa
 couplings, but is not favored for any other reason. The ranges of the \CBLSSM
 parameters for each scan are shown in \TAB~\ref{tab:scanranges}.

The mass parameter ranges were chosen to be consistent with non-observation
 of sparticles at the LEP and LHC experiments\footnote{As discussed in \cite{Krauss:2012ku} the
 bounds on $m_{Z'}$ are about 300 GeV lower than claimed by the LHC experiments once
 gauge kinetic mixing is taken into account  properly. For squarks and gluinos we required them
 to be above one TeV.}, while remaining in a region where
 the LHC may be able to see some phenomena of the model, while the couplings
 were chosen to cover the perturbative range. Thus we generated sets of
 parameter points over the full region of phenomenological interest where \RP
 is possibly conserved.

In general, picking a set of GUT-scale parameters is very unlikely to result in
 a potential which has a phenomenologically acceptable minimum. Instead of
 wasting time searching the parameter space for such points, the strategy
 adopted in \SPheno (and others, such as \texttt{ISAJET} \cite{Paige:2003mg},
 \texttt{SOFTSUSY} \cite{Allanach:2001kg} and
\texttt{SUSPECT} \cite{Djouadi:2002ze}) is to instead fix some parameters
 (in our case, $\mu, B_{\mu}, {\mu}', B_{{\mu}'}$) at the SUSY scale, defined as
 $Q_{SUSY}=\sqrt{m_{\tilde t_1}m_{\tilde t_2}}$, by demanding
 that the tadpole equations are satisfied for \vevs given as input (indirectly,
 by specifying $m_{Z}, \tan\beta, \mZp, \tan{\beta}'$).
 This is sufficient to
 ensure that the chosen point is an extremum, but does not guarantee that the
 extremum is not a saddle point or maximum. Any parameter points that were
 found to have such engineered extrema at their input \vevs which were
 saddle points or maxima (by the presence of tachyons) were discarded.

\begin{table} 
\centering
\begin{tabular}{|c|c|c|} 
\hline
Parameter & \multicolumn{2}{c|}{Common to both} \\ 
\hline 
$M_{1/2}$ & \multicolumn{2}{c|}{\phantom{0}100 \textendash \  1000} \\
$M_{0}$ & \multicolumn{2}{c|}{\phantom{0}100 \textendash \ 3000} \\
$A_{0}$ & \multicolumn{2}{c|}{-3000 \textendash \ 3000\phantom{-}} \\
$\tan\beta$ & \multicolumn{2}{c|}{\phantom{5}3 \textendash \ 45} \\
$m_{Z'}$ & \multicolumn{2}{c|}{1500 \textendash \ 3000} \\
$\tan{\beta}'$ & \multicolumn{2}{c|}{1.0 \textendash \ 1.5} \\
\hline \hline
Parameter & Democratic & Hierarchical \\ 
\hline 
$Y_{x}^{1 1}$ & 0.05 \textendash \ 0.6 & fixed $10^{-3}$ \\
$Y_{x}^{2 2}$ &0.05 \textendash \ 0.6& fixed $10^{-2}$ \\
$Y_{x}^{3 3}$ &0.05 \textendash \ 0.6 & 0.1 \textendash \ 0.6 \\
\hline
\end{tabular} 
\caption{Ranges of parameters used in generating the samples. All dimensionful
 quantities are to be read as in units of GeV. The signs of $\mu$ and $\mu'$
 were fixed to both be positive. The democratic scan consisted of 2330
 points, the hierarchical of 1640 points.} 
\label{tab:scanranges}
\end{table}

\subsection{Finding all the tree-level extrema}

As mentioned in the introduction, finding the global minimum of even the
 tree-level potential as a function of several scalar fields is highly
 non-trivial.
The Gr{\"{o}}bner basis method can be used to find all the minima of a
 polynomial function, as has recently seen some discussion in the
 literature \cite{Maniatis:2006jd,Gray:2008zs}.
 A less well-known method is that of homotopy continuation
 \cite{sommesenumerical, li2003numerical}, which has found use in
 several areas of physics \cite{Mehta:2009zv, Mehta:2011xs, Mehta:2012qr}, in
 particular finding string theory vacua \cite{Mehta:2011wj, Mehta:2012wk} and
 extrema of extended Higgs sectors \cite{Maniatis:2012ex}, where the
 authors investigated a system of two Higgs doublets with up to five singlet
 scalars in a general tree-level potential. In contrast, the Gr{\"{o}}bner basis
 method is deemed prohibitively computationally expensive for systems
 involving more than five or six degrees of freedom \cite{Maniatis:2006jd},
 while for our purposes eight to ten degrees of freedom
 are necessary, and the solutions for thousands of parameter points needed to be
 calculated.

The numerical polyhedral homotopy continuation method is a powerful way to find
 all the roots of a system of polynomial equations quickly \cite{lee2008hom4ps}.
 Essentially it
 works by continuously deforming a simple system of polynomial equations with
 known roots, with as many roots as the classical B{\'{e}}zout bound of the
 system that is to be solved (\IE~the maximum number of roots it could have).
 The simple system with known roots is continuously deformed into the target
 system, with the position of the roots updated with each step. We made use of
 the publically-available program \homps\ \cite{lee2008hom4ps} to
 find all the extrema of the tree-level potential via the polynomial tree-level
 tadpole equations. The tadpole equations for the simple
 case of one generation of sneutrinos with non-zero \vevs are given by
 \EQS~(\ref{eq:tadpole_d})-(\ref{eq:tadpole_a}) below for illustration.
 For each parameter point, the complete set of tadpole equations for the
 appropriate number of
 allowed non-zero sneutrino \vevs were passed to \homps, which provided
 the \vev configurations of all the solutions of the system of equations.

\subsubsection{Tadpole equations}
\label{sec:tadpoles}

If we consider \RPVn in the case where only one generation of
sneutrino acquires a
 \vev and restrict ourselves to real parameters, 
 the tree-level potential minimization conditions, or tadpole equations,
 are:
\begin{align} 
\label{eq:tadpole_d}
t_d =& \frac{1}{8} v_d \Big(\bar{g} g_{BL} \Big(-2 v_{\bar{\eta}}^{2}  + 2 v_{\eta}^{2}  - v_{R}^{2}  + v_{L}^{2}\Big) + \Big(g_{1}^{2} + \bar{g}^{2} + g_{2}^{2}\Big)\Big(- v_{u}^{2}  + v_{d}^{2} + v_{L}^{2}\Big)\Big)\nonumber \\ 
 &- \frac{1}{\sqrt{2}} v_L v_R Y_\nu \mu +v_d \Big(m_{H_d}^2 + \mu^{2}\Big)- v_u B_{\mu} \\ 
t_u =& \frac{1}{8} v_u \Big(\Big(- \bar{g}^{2}  - g_{1}^{2}  - g_{2}^{2} \Big)\Big(- v_{u}^{2}  + v_{d}^{2} + v_{L}^{2}\Big) + \bar{g} g_{BL} \Big(2 v_{\bar{\eta}}^{2}  -2 v_{\eta}^{2}  - v_{L}^{2}  + v_{R}^{2}\Big)\Big)\nonumber \\ 
 &+\frac{1}{2} \Big(-2 v_d B_{\mu}  + v_L v_R \Big(2 v_{\eta} Y_x Y_\nu  + \sqrt{2} T_\nu \Big) + v_u \Big(2 \Big(m_{H_u}^2 + \mu^{2}\Big) + \Big(v_{L}^{2} + v_{R}^{2}\Big)Y_{\nu}^{2} \Big)\Big)\\ 
t_L =& \frac{1}{8} v_L \Big(\Big(g_{1}^{2} + \bar{g}^{2} + g_{2}^{2}\Big)\Big(- v_{u}^{2}  + v_{d}^{2} + v_{L}^{2}\Big)+g_{BL}^{2} \Big(-2 v_{\bar{\eta}}^{2}  + 2 v_{\eta}^{2}  - v_{R}^{2}  + v_{L}^{2}\Big)\nonumber \\ 
 &+\bar{g} g_{BL} \Big(-2 v_{\bar{\eta}}^{2}  + 2 \Big(v_{L}^{2} + v_{\eta}^{2}\Big) - v_{R}^{2}  - v_{u}^{2}  + v_{d}^{2}\Big)\Big)\nonumber \\ 
 &+\frac{1}{2} \Big(2 m_L^2 v_L  + \sqrt{2} v_R v_u T_\nu  + Y_\nu \Big(v_L \Big(v_{R}^{2} + v_{u}^{2}\Big)Y_\nu  + v_R \Big(2 v_u v_{\eta} Y_x  - \sqrt{2} v_d \mu \Big)\Big)\Big)\\ 
t_R =& \frac{1}{8} g_{BL} v_R \Big(- \bar{g} \Big(- v_{u}^{2}  + v_{d}^{2} + v_{L}^{2}\Big) + g_{BL} \Big(2 v_{\bar{\eta}}^{2}  -2 v_{\eta}^{2}  - v_{L}^{2}  + v_{R}^{2}\Big)\Big)\nonumber \\ 
 &+\frac{1}{2} \Big(2 v_{R}^{3} Y_{x}^{2} +v_R \Big(-2 \sqrt{2} {\mu'} v_{\bar{\eta}} Y_x  + 2 \Big(\sqrt{2} v_{\eta} T_x  + {m_{\nu^c}^2}\Big) + 4 v_{\eta}^{2} Y_{x}^{2}  + \Big(v_{L}^{2} + v_{u}^{2}\Big)Y_{\nu}^{2} \Big)\nonumber \\ 
 &+v_L \Big(\sqrt{2} v_u T_\nu  + Y_\nu \Big(2 v_u v_{\eta} Y_x  - \sqrt{2} v_d \mu \Big)\Big)\Big)\\ 
t_{\eta} =& \frac{1}{4} g_{BL} v_{\eta} \Big(\bar{g} \Big(- v_{u}^{2}  + v_{d}^{2} + v_{L}^{2}\Big) + g_{BL} \Big(-2 v_{\bar{\eta}}^{2}  + 2 v_{\eta}^{2}  - v_{R}^{2}  + v_{L}^{2}\Big)\Big)+v_{\eta} \Big(2 v_{R}^{2} Y_{x}^{2}  + m_{\eta}^2 + {\mu'}^{2}\Big)\nonumber \\ 
 &+v_L v_R v_u Y_x Y_\nu - v_{\bar{\eta}} {B_{\mu'}} +\frac{1}{\sqrt{2}} v_{R}^{2} T_x \\ 
t_{\bar{\eta}} =& \frac{1}{4} g_{BL} v_{\bar{\eta}} \Big(- \bar{g} \Big(- v_{u}^{2}  + v_{d}^{2} + v_{L}^{2}\Big) + g_{BL} \Big(2 v_{\bar{\eta}}^{2}  -2 v_{\eta}^{2}  - v_{L}^{2}  + v_{R}^{2}\Big)\Big) \nonumber \\
& - \frac{1}{\sqrt{2}} {\mu'} v_{R}^{2} Y_x  + \Big(m_{\bar{\eta}}^2 + {\mu'}^{2}\Big)v_{\bar{\eta}}  - v_{\eta} {B_{\mu'}} 
\label{eq:tadpole_a}
\end{align} 
Obviously this is a highly coupled system of cubic equations in the \vevs and 
can only be solved by numerical methods. 
One might be tempted to simplify these
equations by using 
$|v_L/v_d|, (v^2_d+v^2_u)/(v_\eta^2+v_{\bar{\eta}}^2+v_R^2) \ll 1$ 
to satisfy 
the phenomenological constraints such as the required hierarchy of the vector
 boson masses or neutrino masses. However, for our purposes we are only allowed
 to do so to engineer the starting minimum, but have to use the general formulas 
 to check if the phenomenlogically allowed minimum is indeed the deepest one.

\subsubsection{Vacuum expectation value phase rotations}

Degenerate lines of minima, corresponding to
 unconstrained phases of some of the \vevs, cause problems since the homotopy
 continuation method maps discrete numbers of solutions of a simple system to
 discrete numbers of solutions of the target system. From the physics
point of view this is equivalent to the statement that only phase differences
are relevant but not the phase of a single parameter itself. Already at the
 level of the superpotential in \EQ~(\ref{eq:superpot}) it is easy to see that
 two \vevs in the extended Higgs sector
can be chosen real without changing the phase of a single parameter.
We chose to take the \vevs of $H_{u}$ and $\bar{\eta}$
 as purely real, and all other phases to be relative to these\footnote{One 
 could in principle also take in
 addition the \vevs of  of $H_{d}$ and $\eta$ to be real be redefining the
 phases of
$\mu$ and $\mu'$. However, as this cannot be done consistently within \homps
we take them to be complex and keep the two parameters real.}.
As we are neglecting the flavour structure of $Y_\nu$, we can also choose 
all $v_{R,i}$ to be real (or equivalently we could take
all $v_{L,i}$ real). If this flavour structure is taken into account,
only one of the six sneutrino \vevs can be taken real in general. 
However, even assuming that we can neglect flavour mixing in $Y_\nu$, we
have five real and five complex \vevs, resulting in fifteen real parameters
which have to be calculated by \homps. Even though it is relatively 
fast, solving a system with ten degrees
 of freedom in typically twenty minutes on a single 3.4 GHz Intel i7 core,
it takes about 10 days for the 15 degrees of freedom. For this reason
we decided that for the democratic scan, all the \vevs
 would be taken to be real.
 Three points out of the 2330 were found which had one
 or more mass-squared that was negative at the deepest extremum found, and these
 are taken to have deeper minima where at least one of the  \vevs is complex.
In all three cases, at least one sneutrino \vev was non-zero. 
The hierarchical case can effectively be treated like a one generation model
and here we allowed for complex \vevs.
We found in this case that every minimum with a non-zero imaginary part for any
 \vev was degenerate with another minimum of the same parameter point with
 purely real \vevs, with the same magnitudes but some different
 signs.
 Hence our conclusions would be unchanged if we had restricted ourselves
 to purely
 real \vevs in the hierarchical scan.

\subsection{Loop corrections}

Before reaching any conclusion we have to consider the loop
 corrections to the scalar potential which are known to be important. 
 The homotopy continuation method allows
 us to find all the solutions to the tree-level minimization conditions. However
 the effective potential at the one-loop level
 already has a more complicated structure including logarithmic
 contributions. 

Loop corrections might modify our previous discussion by changing the relative
 depth between \RPCg and \RPVg minima or promoting a gauge-conserving minimum to
 be the global minimum at loop level. It might also happen (although less
 likely) that that the nature of the extremum is changed, \EG~a tree-level
 saddle point becomes a lower-lying minimum of the effective potential. 

In order to assess the effect of loop corrections, we evaluated the one-loop
 effective potential at all the extrema found by
 \homps for every parameter point. In its more general form, the expression for
 the effective potential contains field-dependent masses. As our study centers
 around minima for the potential, the field-dependent masses are just the
 tree-level masses as function of the \vevs. We chose to remain in the Landau
 gauge and $\overline{\text{DR}}'$ scheme and follow the results
 in \cite{Martin:2002}. 

Specifically, to evaluate the effective potential for a given set of \vevs we
 used the
 expression:
\begin{equation}
  V^1 = V^0
  + \frac{1}{16 {\pi}^2} \sum_n \left( -1 \right)^{2S_n}\left( 2 S_n
 + 1 \right) \left( \frac{m_n^4}{4} \left[ \ln\left(\frac{ m_n^2}{ Q^2} \right)
  - \frac{3}{2} \right] \right),
\label{eq:EffPot}
\end{equation}
where $V^0$ is the tree-level scalar potential with corrections of the same
 polynomial form from the finite parts of the counterterms and $n$ goes over
 all the real
 scalar ($S_n=0$), Weyl fermion ($S_n=\frac{1}{2}$) and vector ($S_n=1$) degrees
 of freedom.
 The $m_n$ are the tree-level masses for a given set
 of parameters and \vevs at the renormalization scale
 $Q = \sqrt{m_{\tilde t_1}m_{\tilde t_2}}$. The renormalization scale
 is fixed by the stop masses for the \vev configuration given as input, and is
 held fixed for comparison with other \vev configurations, as changing the scale
 in evaluating the one-loop potential for different \vev configurations is
 inconsistent for a fixed-order calculation \cite{Ferreira:2000hg}.
A more detailed formula is given in appendix \ref{app:effpot}.

The masses were evaluated by a routine taking the tree-level expressions
 from \SARAH, including the \RPVg mixings of neutrinos with neutralinos, charginos with charged leptons, 
 charged Higgs bosons with charged sleptons and
 sneutrinos with neutral Higgs together with bilepton scalars/pseudoscalars in the
 presence of sneutrino \vevs. The corresponding formulas are given in
 appendix \ref{app:massmatrices}. These masses were then used to calculate
 the full effective one-loop potential according to \EQ~(\ref{eq:EffPot}).

When the potential is non-convex the conventional loop expansion becomes complex
 \cite{Fujimoto:1982tc}.
Specifically, the value of the potential on a saddle point can be interpreted as
 an amplitude with a corresponding decay width (its complex part)
\cite{Weinberg:1987}. It was sufficient to use
 the real part of the effective potential to estimate whether a saddle point
 might have been promoted to a lower-lying minimum by comparing it with the
 value of the effective potential in all of the other minima. 
     
By doing this it was possible to determine whether the hierarchy for the minima
 found at tree level was respected once loop corrections are included. We found
 that loop corrections have a non-negligible impact to the tree-level
 conclusions.

When this analysis indicated that a parameter point had a deeper minimum
 than the input minimum, we ran the code
 \texttt{Vevacious} \cite{Vevacious} with a model file generated by \SARAH to
 i) check that the one-loop potential did indeed have a minimum near this \vev
 configuration and that it was indeed deeper, using \texttt{MINUIT}
 \cite{James:1975dr}, and ii) estimate the tunneling
 time to the deeper minimum, using \texttt{CosmoTransitions}
 \cite{Wainwright:2011kj}.

\section{Results}
\label{sec:results}

In general, each parameter point of our scans had several minima, both \RPCg and
 \RPVg. We categorized the points by the nature of the lowest of their minima
 as in \TAB~\ref{tab:categorization}, and we present a breakdown of the number of points 
 in each category in \TAB~\ref{tab:category_populations}.
 Every point fell into one of these three
 categories: \EG~there were no points which had a global minimum with no
 unbroken symmetries, and there were no points which broke \RP without breaking
 both \SUL and \UBL. 
As it happens, there were no ``RPC'' parameter points in either scan at either
 tree or one-loop level which had global minima deeper than the input minima, by
 which we mean the set of \vevs chosen as input for \SPheno.

The majority of ``RPV'' points broke \RP through non-zero sneutrino \vevs, but
 there were six points in the hierarchical scan which had one-loop global
 minima with negative stop mass-squared, hence we assume that
 such parameter
 points have non-zero stop \vevs. We plot them along with the other ``RPV''
 points since they do break \RP, in addition to $SU(3)_{c}$ and $U(1)_{em}$.

The third category, ``unbroken'', only appears in the one-loop-level analysis.
 It is known that particularly precarious arrangements of parameters that break
 symmetries spontaneously at tree level may not break these symmetries when
 evaluated at one-loop level \cite{deCarlos1993320}. In such cases, the (not
 obvious) combination of Lagrangian parameters that parameterizes \UBL-breaking
 for example may be of the same magnitude as the loop corrections. We note that
 all such points that we found had zero sneutrino \vevs.

\begin{table} 
\centering
\begin{tabular}{|c|c|} 
\hline
Categorization & Description \\ 
\hline
``RPC''      & \SUL, \UBL both broken, \RP conserved. \\
\hline
``RPV''      & \SUL, \UBL both broken, \RP broken. \\
\hline
``unbroken'' & Either \SUL or \UBL broken but not both, \RP conserved. \\
\hline
\end{tabular} 
\caption{Categorization of parameter points according to the symmetries broken
 by their global minima.}
\label{tab:categorization}
\end{table}

\begin{table} 
\centering
\begin{tabular}{|c|c|c|c|c|} 
\hline
Categorization & \multicolumn{2}{c|}{Hierarchical scan}
                 & \multicolumn{2}{c|}{Democratic scan} \\
\hline
total & \multicolumn{2}{c|}{1640} & \multicolumn{2}{c|}{2330}\\
\hline
 & tree level & one-loop level & tree level & one-loop level \\
\hline
``RPC''      & 1422 & 1275 & 2236 & 2167 \\
\hline
``RPV''      &  218 &  212 &  94 &  86 \\
\hline
``unbroken'' &    0 &  153 &    0 &  77 \\
\hline
\end{tabular} 
\caption{Number of parameter points in the various categories. All of the
 parameter points from both scans categorized as ``unbroken'' broke
 \SUL without breaking \UBL. Not all parameter points that are ``RPC'' at the
 one-loop level were ``RPC'' at tree level, and likewise for the ``RPV''
 category.}
\label{tab:category_populations}
\end{table}

\begin{figure}[hbt]
\begin{minipage}{\linewidth}
\includegraphics[width=0.4\linewidth
               ]{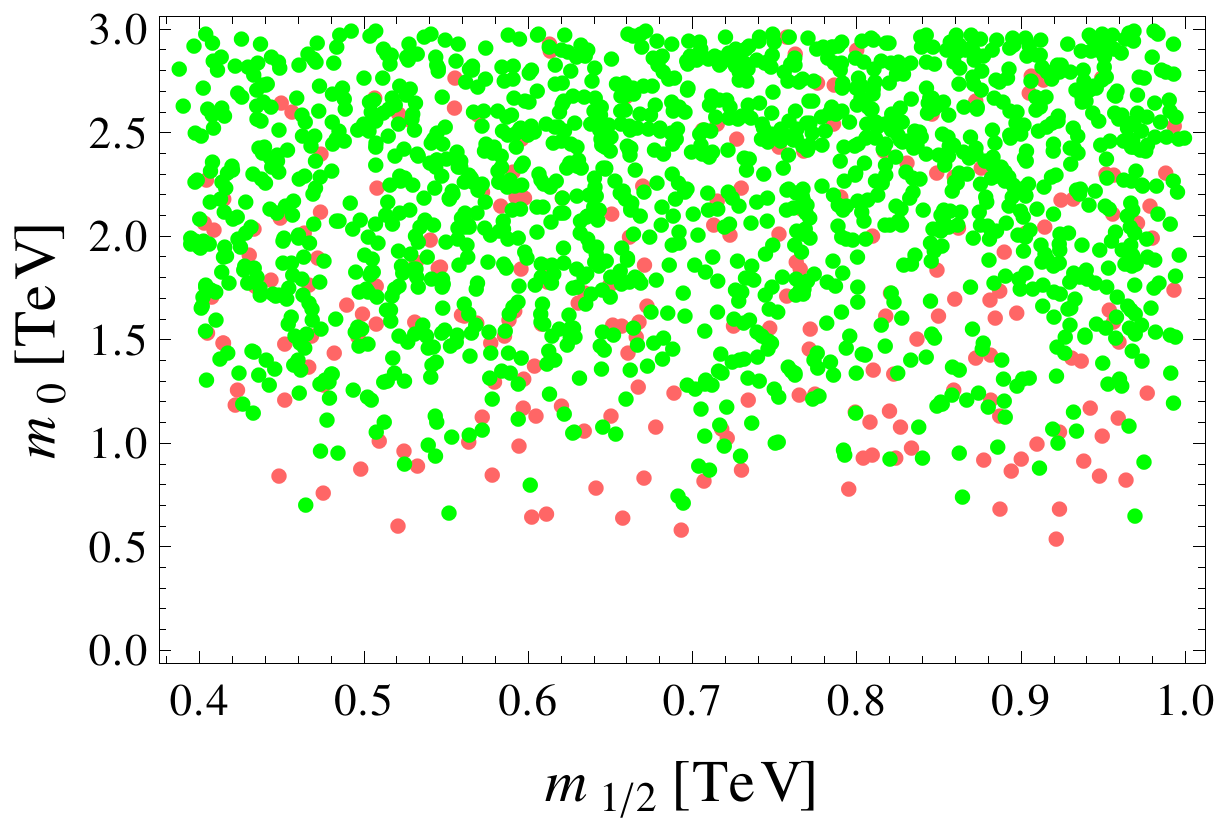}  
\hfill
\includegraphics[width=0.4\linewidth
               ]{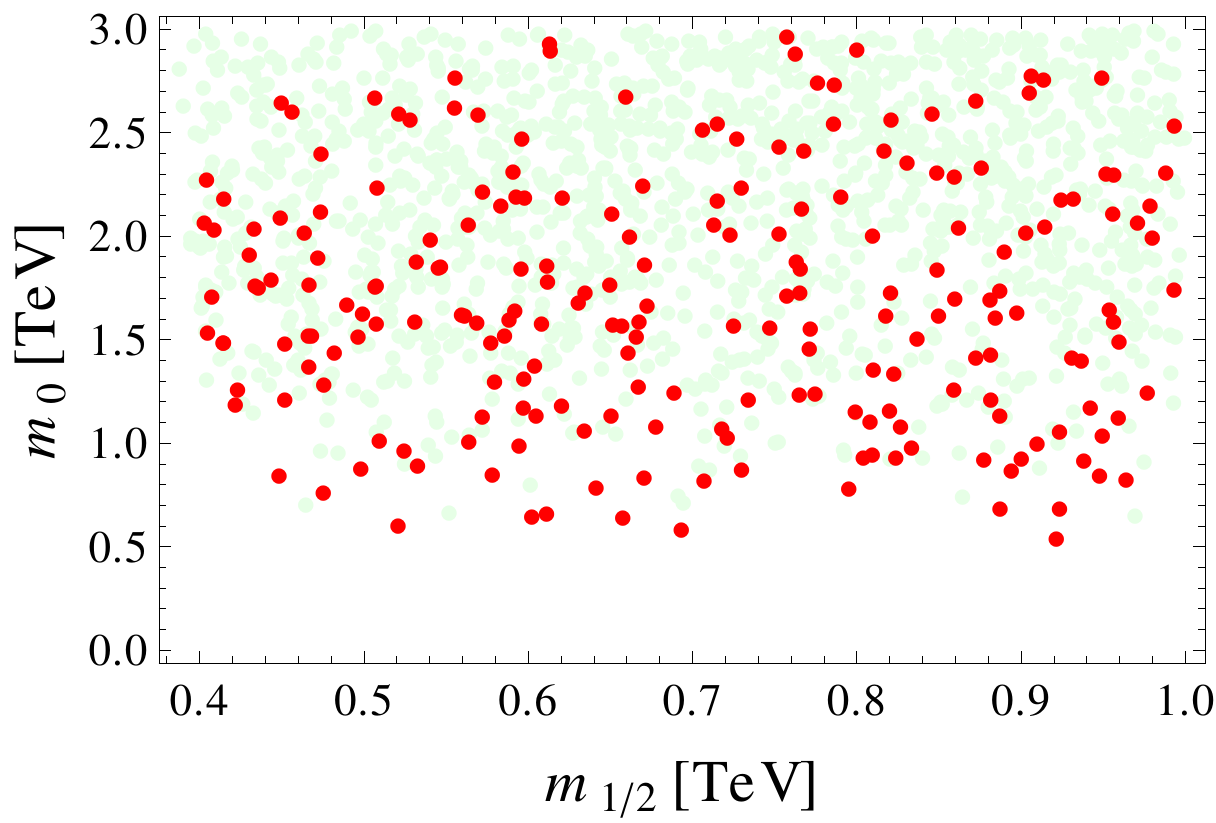}  \\
\vspace{4mm}
\includegraphics[width=0.4\linewidth
               ]{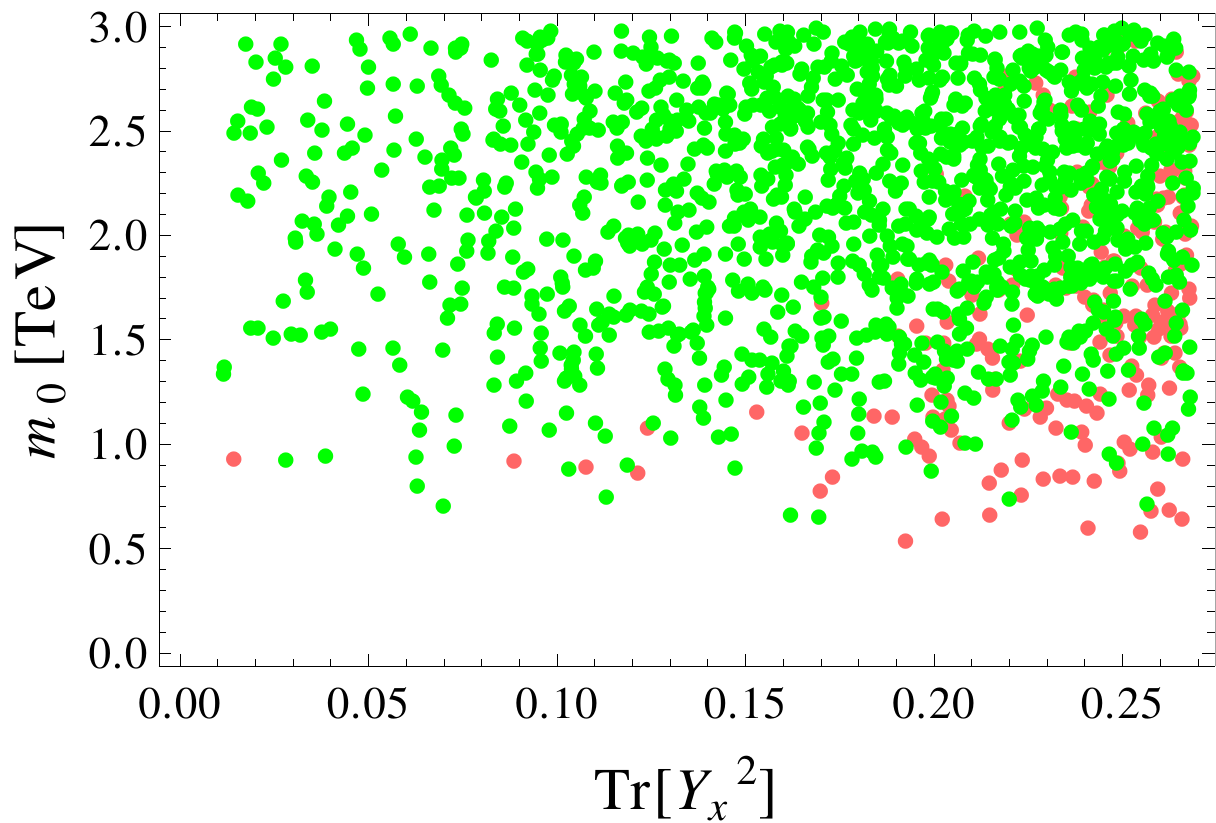}  
\hfill
\includegraphics[width=0.4\linewidth
               ]{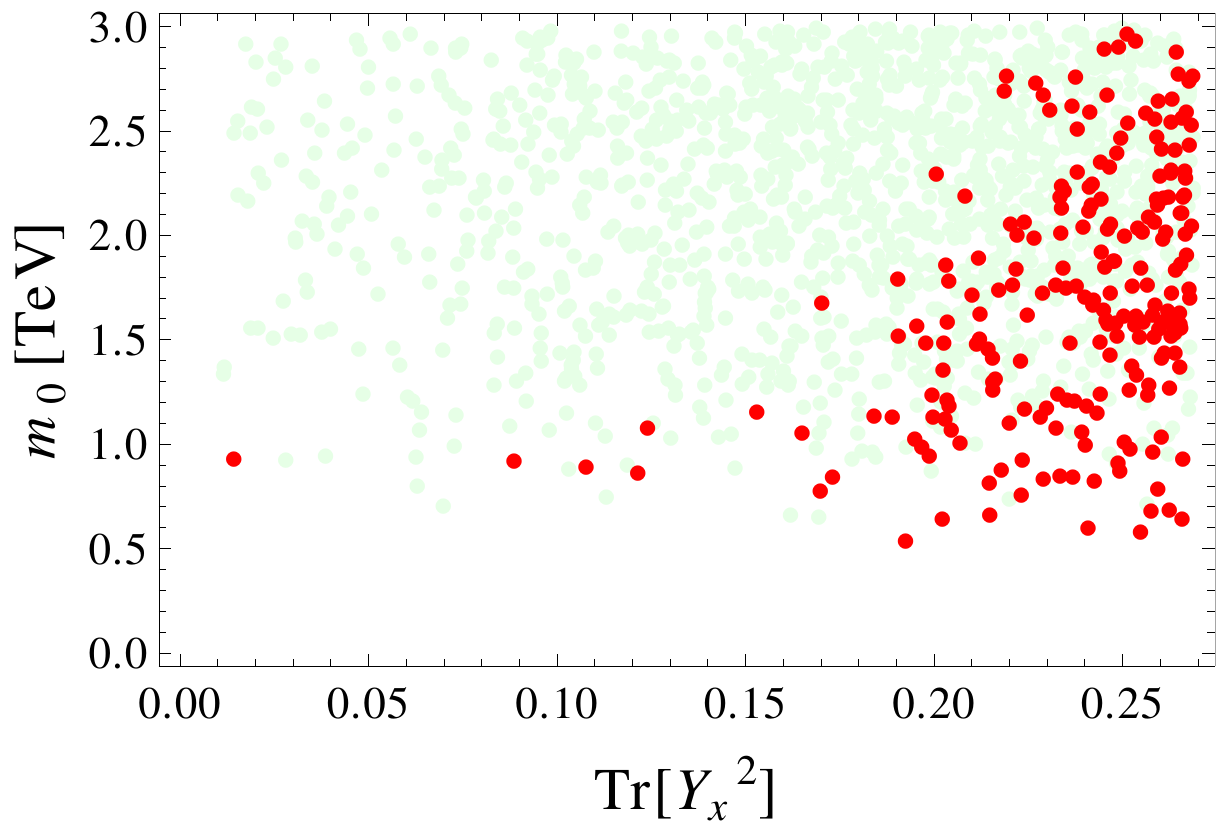}  \\
\vspace{4mm}
\includegraphics[width=0.4\linewidth
               ]{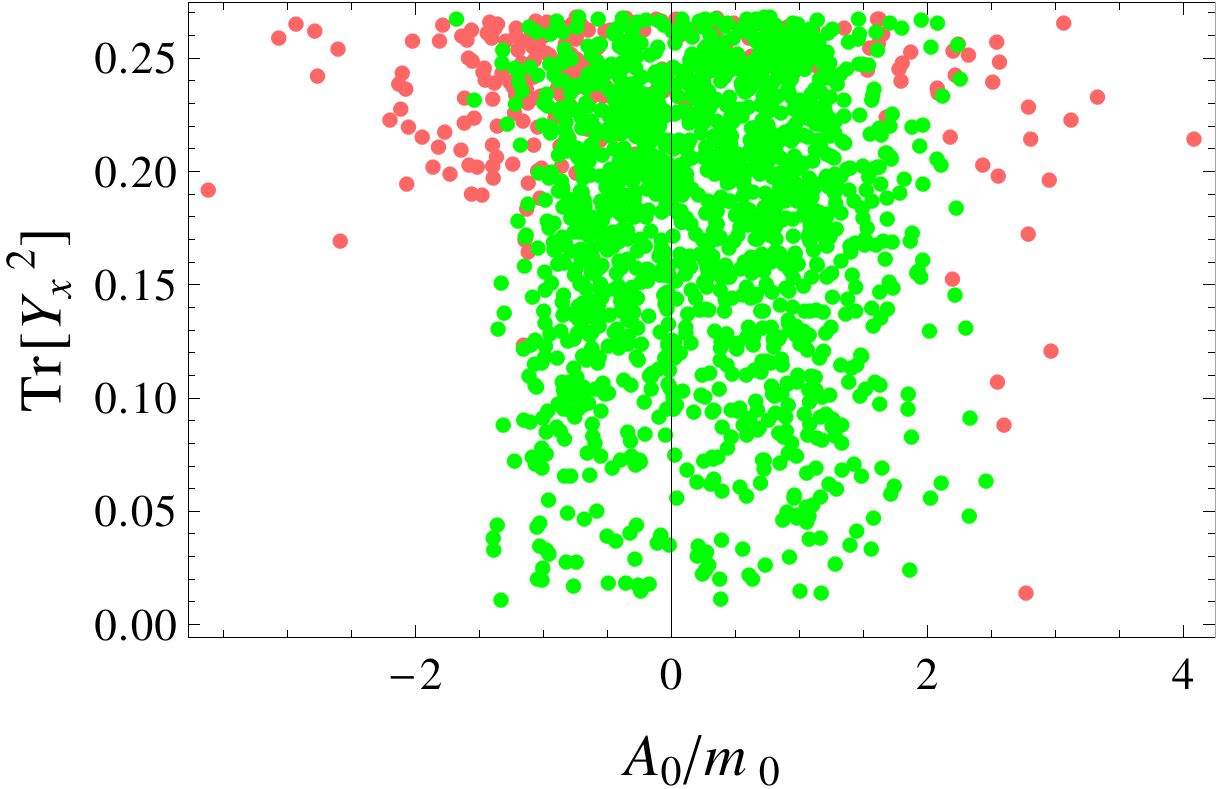}  
\hfill
\includegraphics[width=0.4\linewidth
               ]{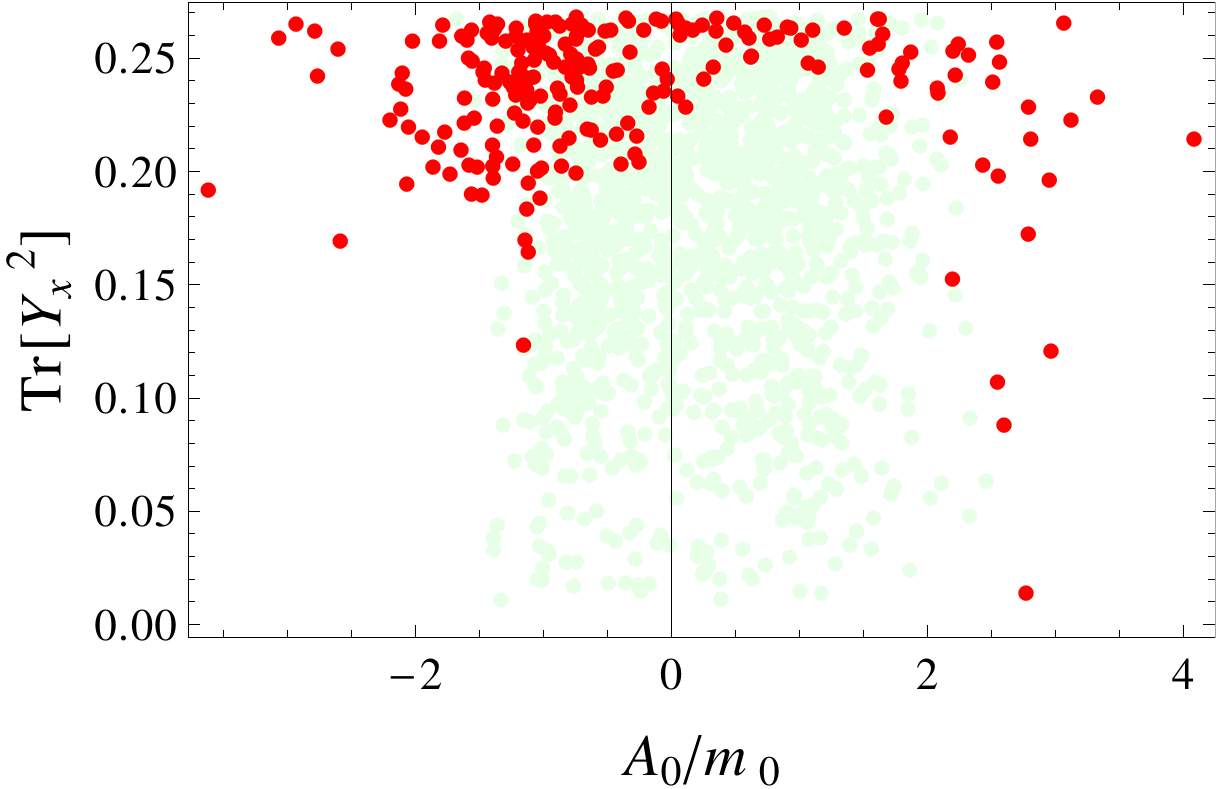}  \\
\vspace{4mm}
\includegraphics[width=0.4\linewidth
               ]{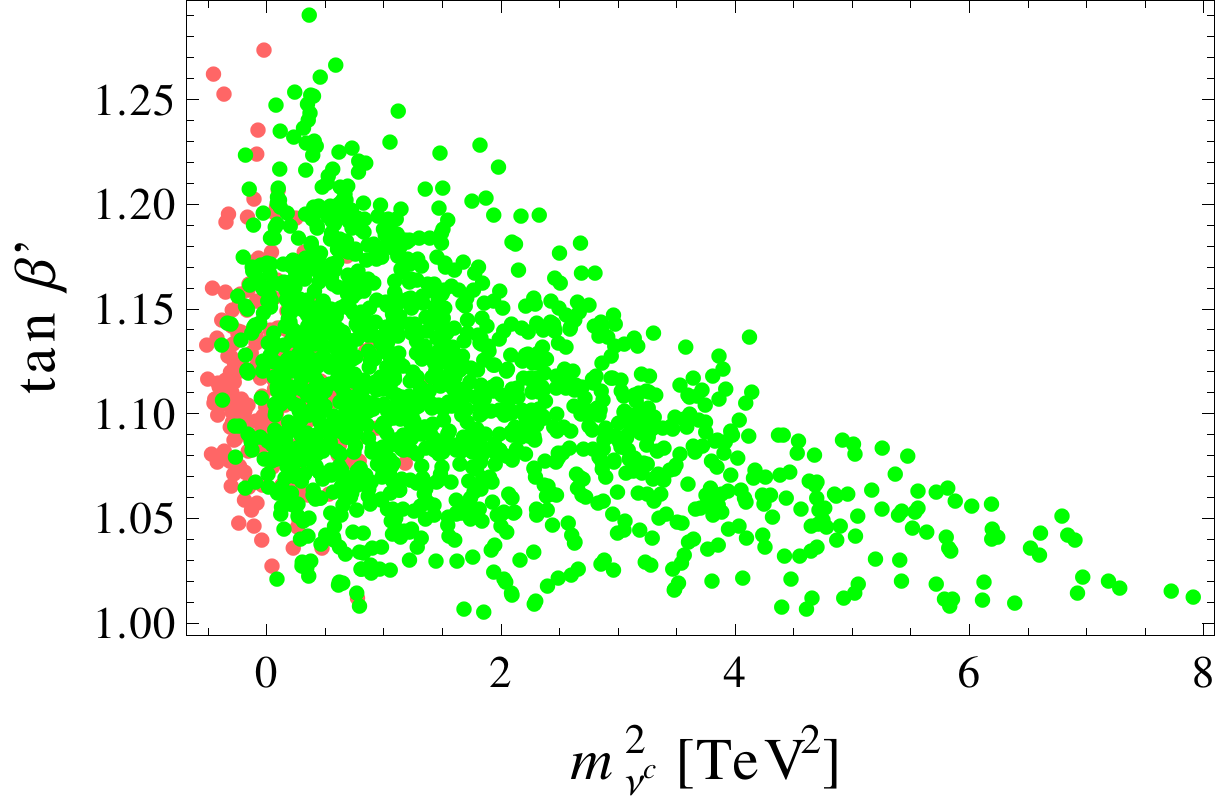}  
\hfill
\includegraphics[width=0.4\linewidth
               ]{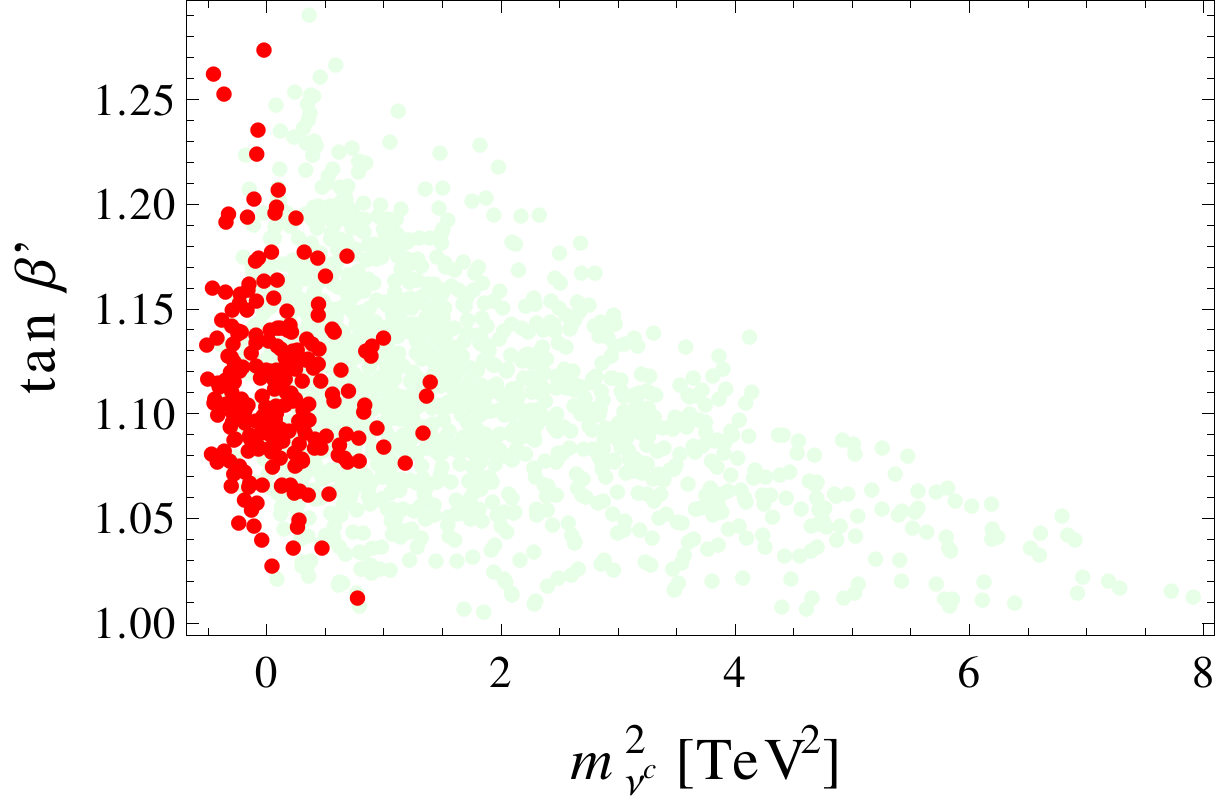}
  \end{minipage}
\caption{Projections into various parameter planes of the 1640 hierarchical scan
 parameter points, categorized by the nature of their global minima (see
 \TAB~\ref{tab:categorization}) at tree level. ``RPC'' points are plotted in
 green (light grey), and ``RPV'' points are plotted in red (medium grey). In the
 plots on the left, the ``RPC'' points are plotted on top of the ``RPV'' points,
 which are faded (very light grey), while in the plots on the right the ``RPV''
 points are plotted on top of the ``RPC'' points, which are faded (very light
 grey). ($\mRSq$ is the lowest or most negative of the three soft
 SUSY-breaking mass-squared parameters for the \rsnus, evaluated at the SUSY
 scale.)}
\label{fig:hier_tree_comparisons}
\end{figure}

\begin{figure}[hbt]
\begin{minipage}{\linewidth}
\includegraphics[width=0.4\linewidth
               ]{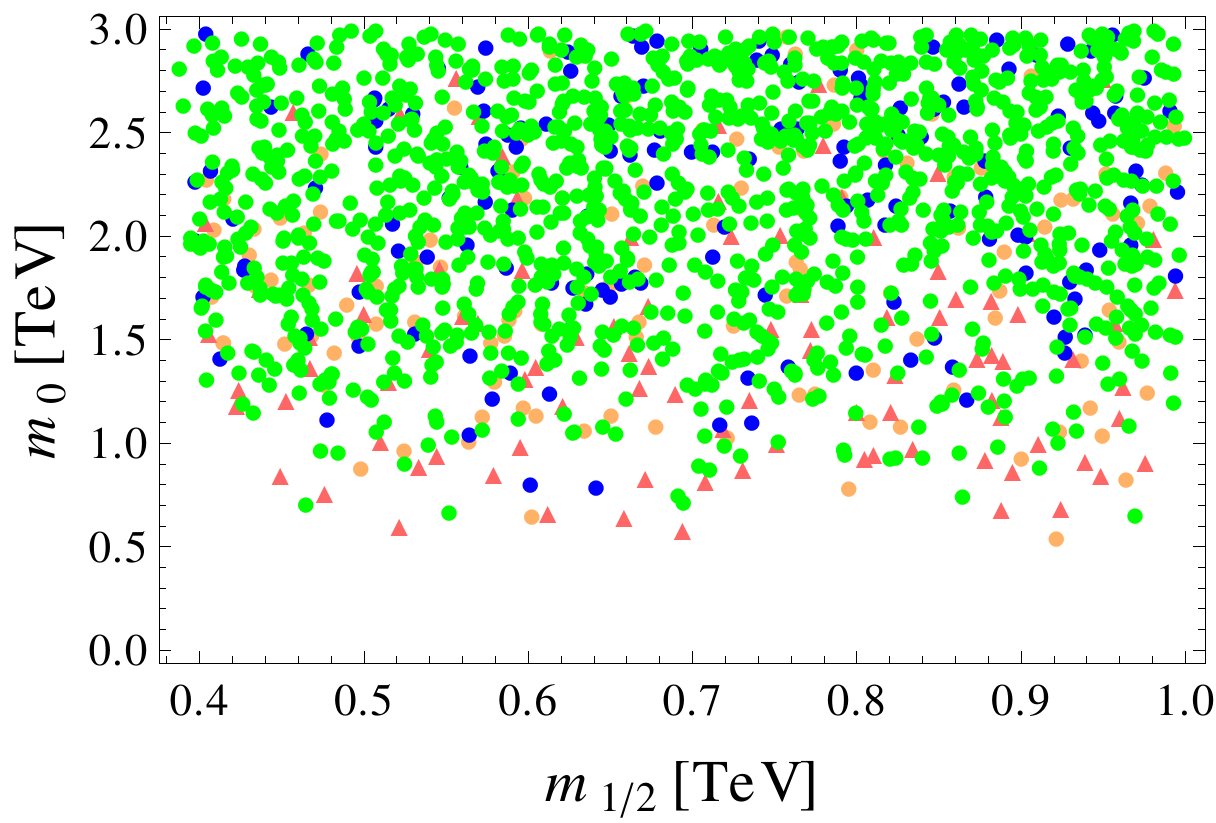}  
\hfill
\includegraphics[width=0.4\linewidth
               ]{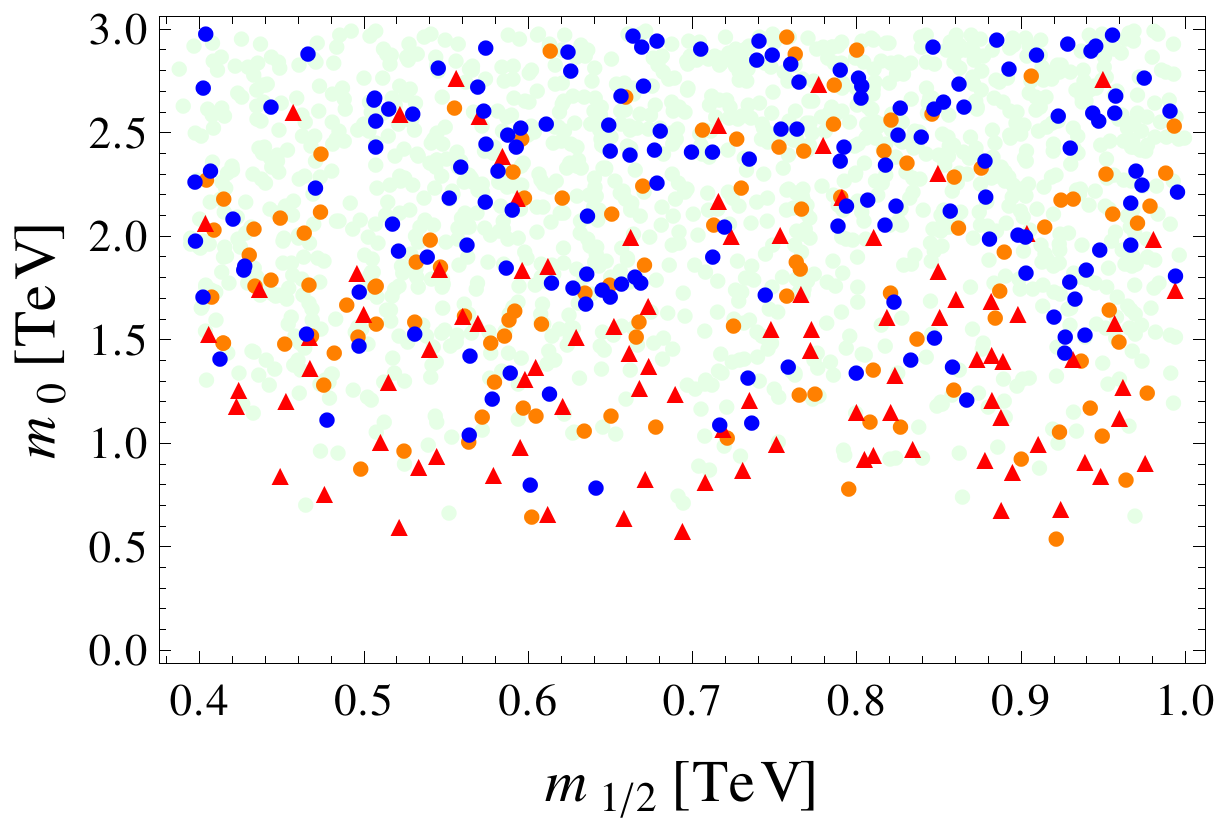}  \\
\vspace{4mm}
\includegraphics[width=0.4\linewidth
               ]{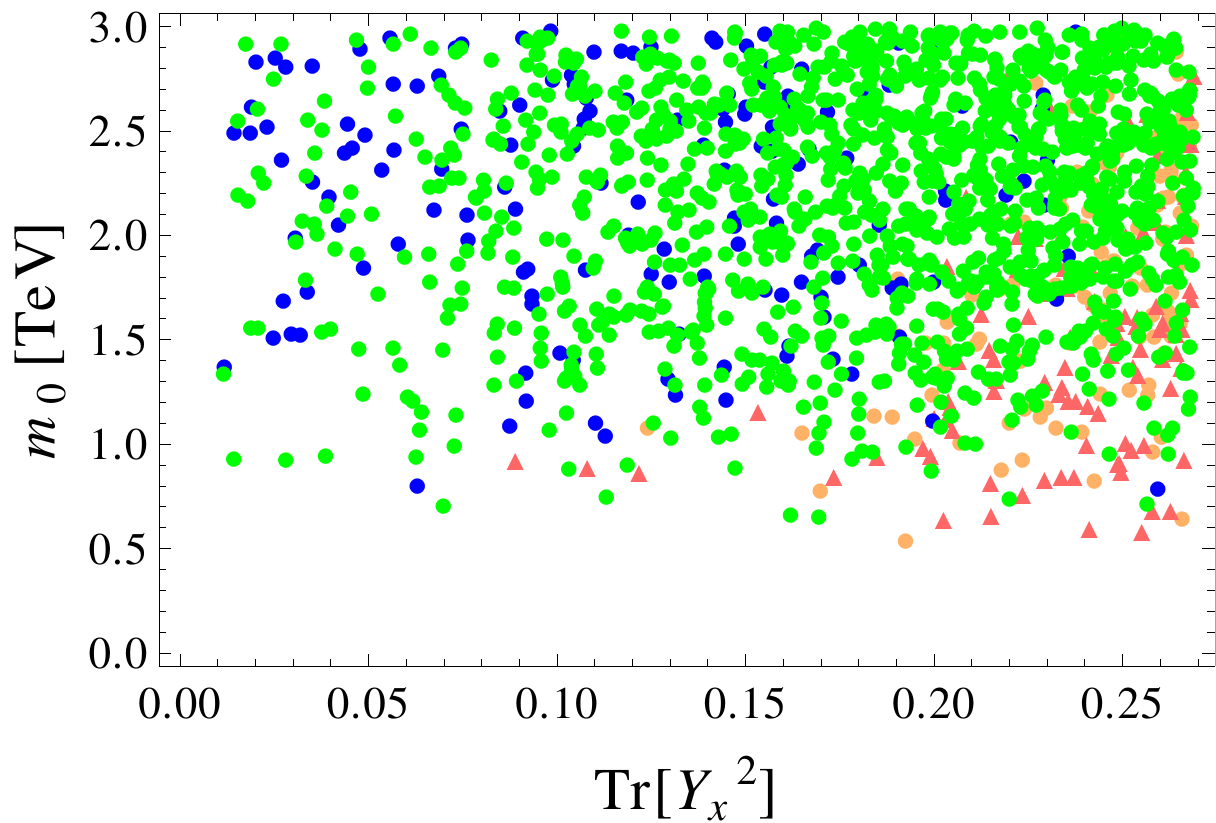}  
\hfill
\includegraphics[width=0.4\linewidth
               ]{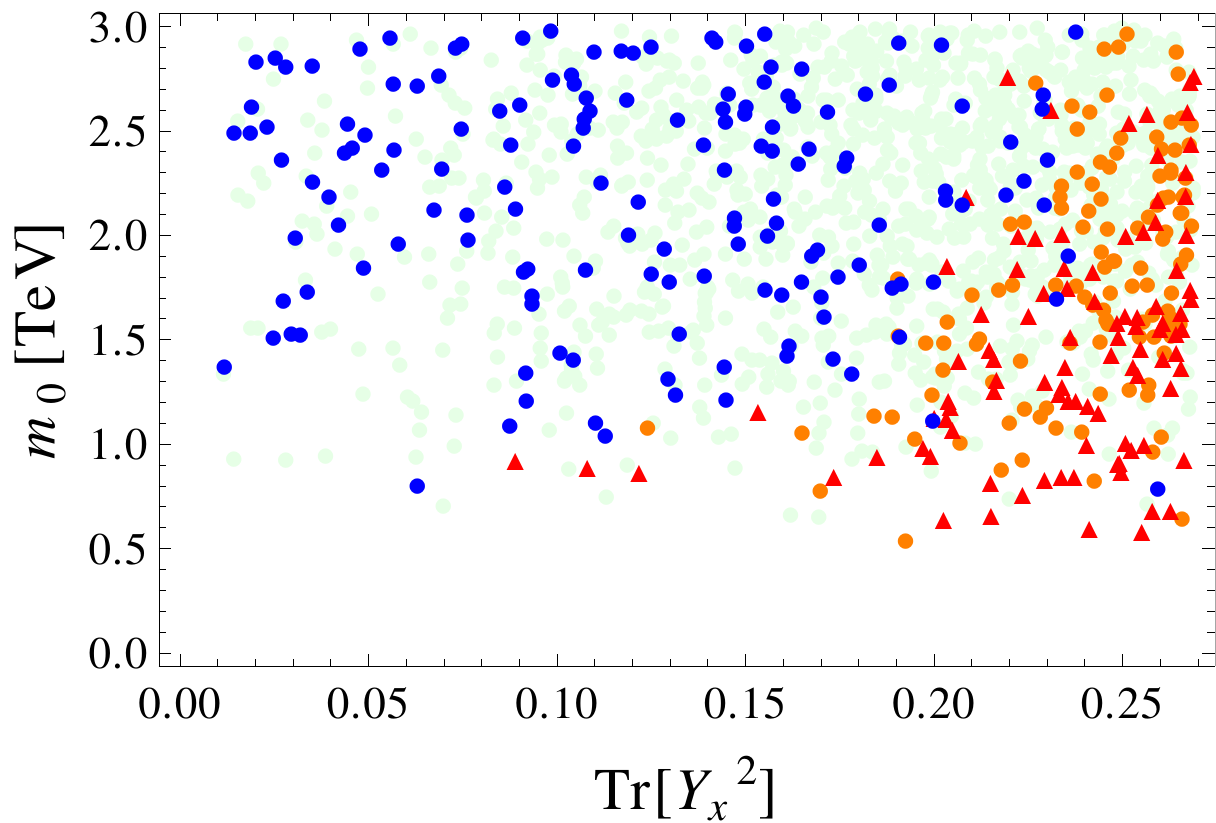}  \\
\vspace{4mm}
\includegraphics[width=0.4\linewidth
               ]{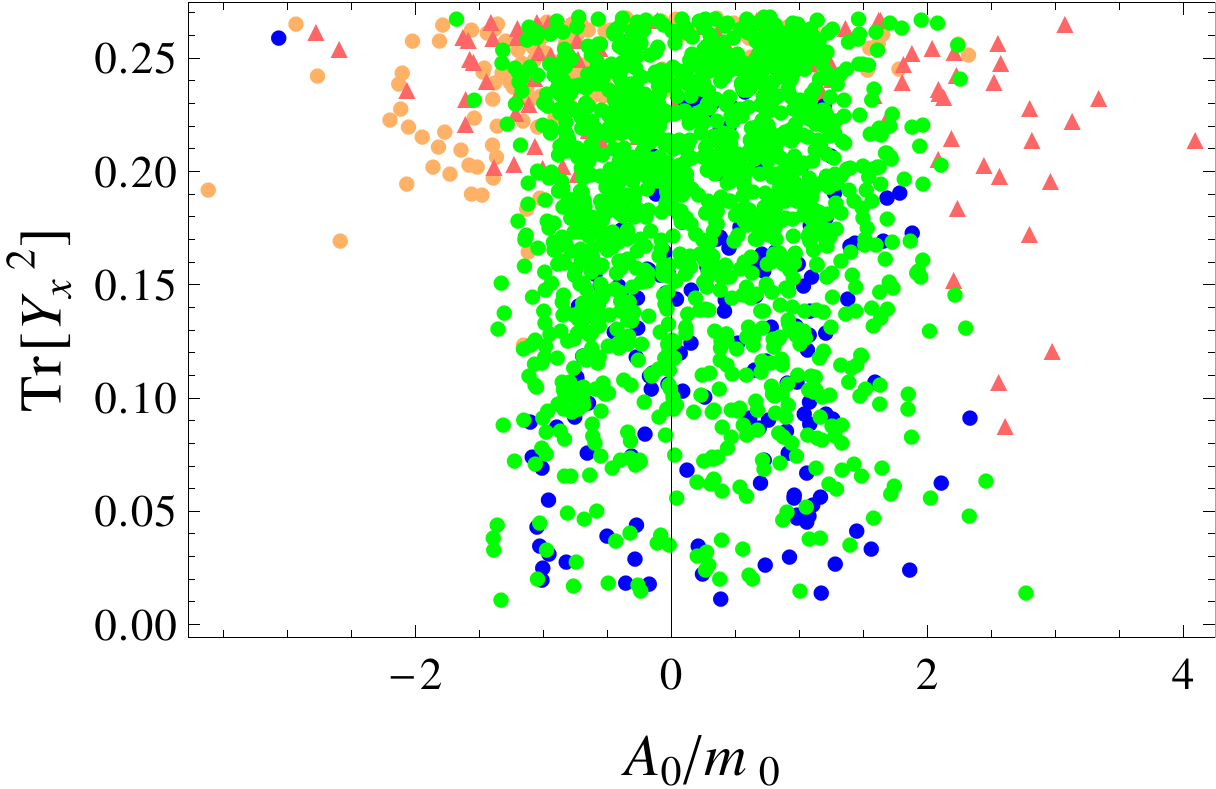}  
\hfill
\includegraphics[width=0.4\linewidth
               ]{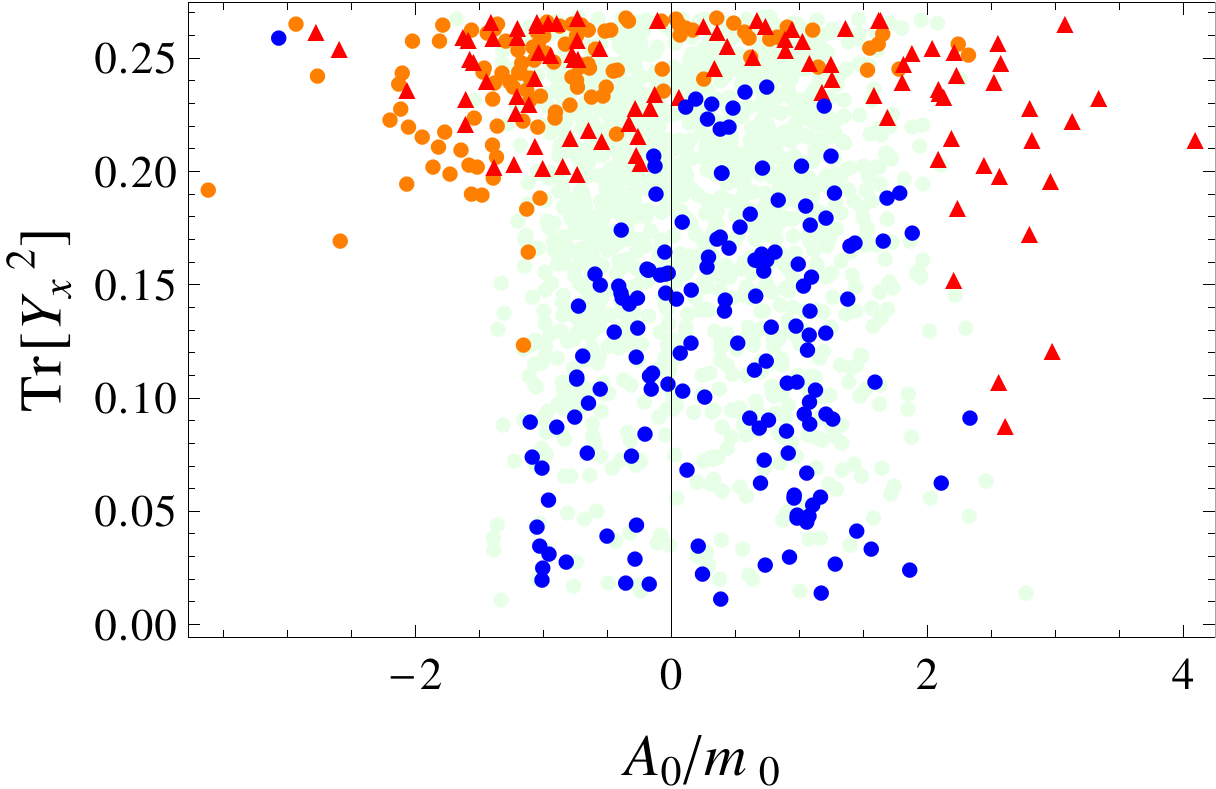}  \\
\vspace{4mm}
\includegraphics[width=0.4\linewidth
               ]{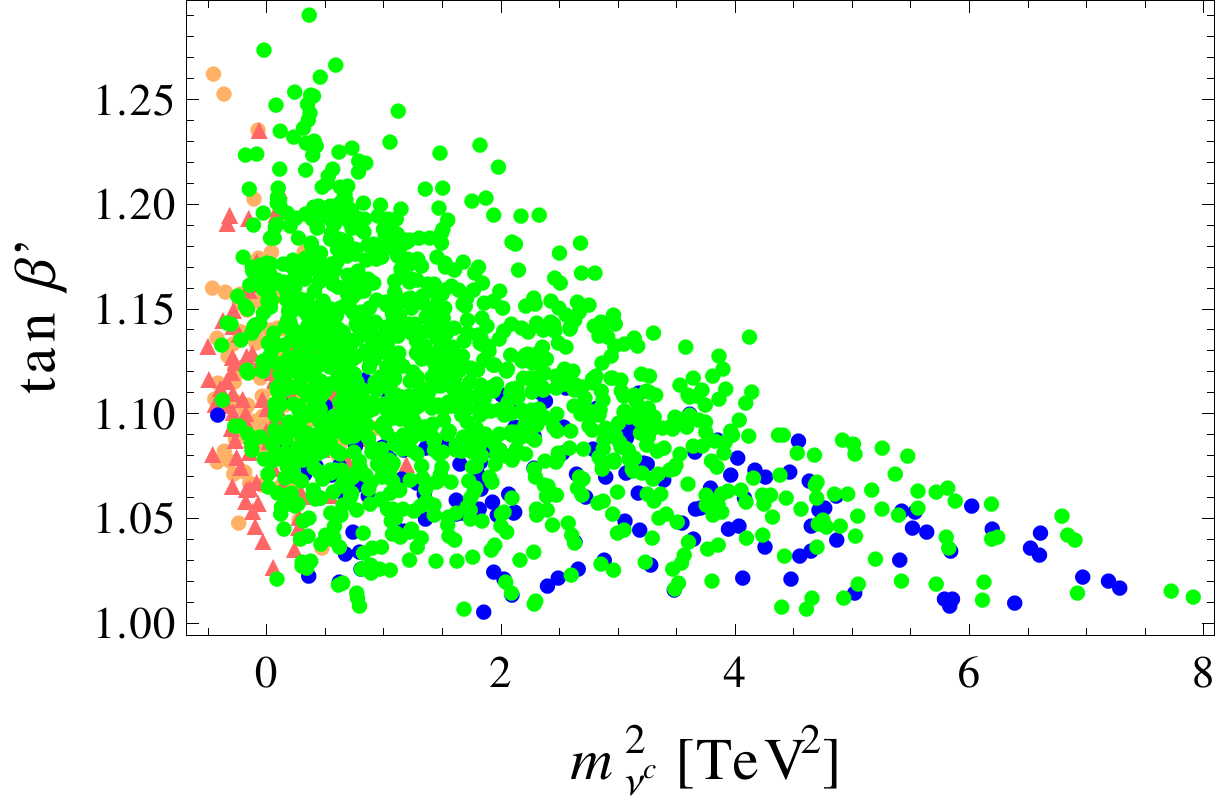}  
\hfill
\includegraphics[width=0.4\linewidth
               ]{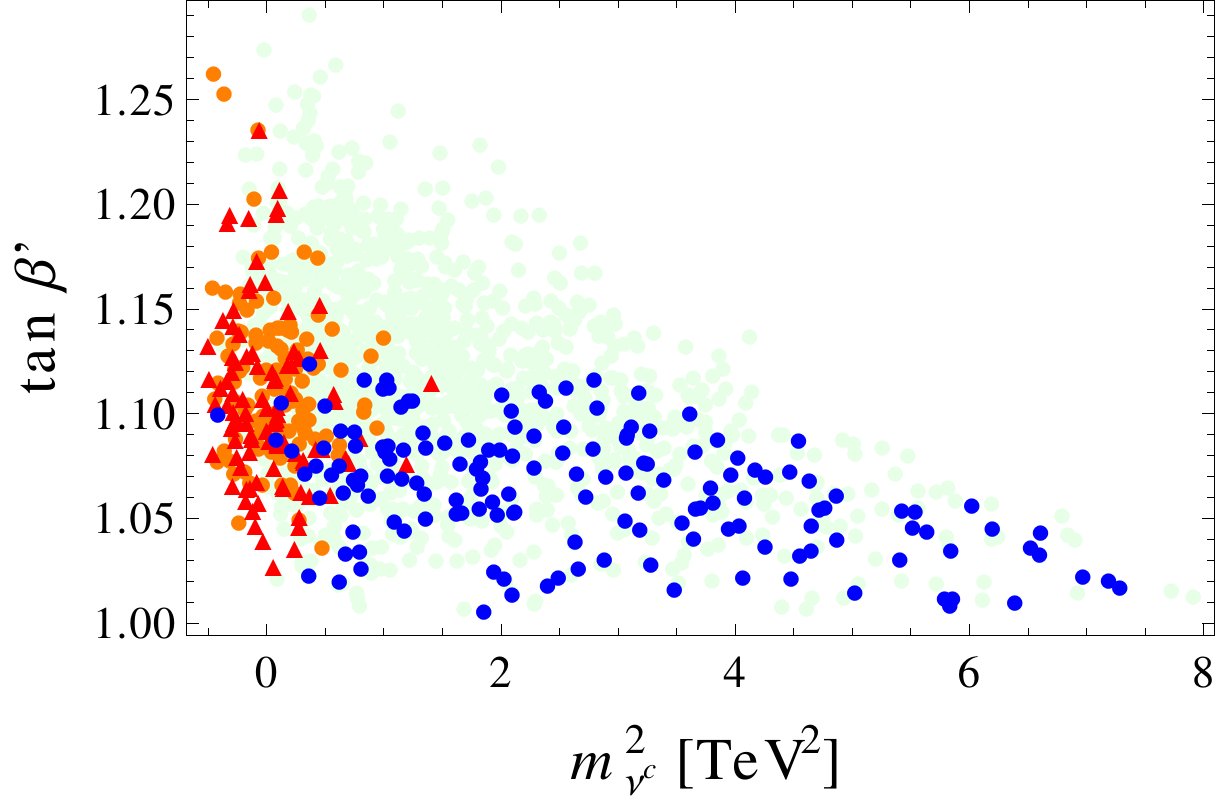}
  \end{minipage}
\caption{Projections into various parameter planes of the 1640 hierarchical scan
 parameter points, categorized by the nature of their global minima (see
 \TAB~\ref{tab:categorization}) at one-loop level.
 As in \FIG~\ref{fig:hier_tree_comparisons}
 ``RPC'' points are plotted in green (light grey), but now the ``RPV'' points
 are in two groups based on the tunneling time from the
 ``RPC'' input minimum to the deeper ``RPV'' minimum: lower than one tenth of
 the age of the Universe as orange (medium-light grey) circles, greater than a
 tenth of the age of the Universe as red (medium grey) triangles.
 ``Gauge conserving'' points are in blue (dark grey).}
\label{fig:hier_loop_comparisons}
\end{figure}

\begin{figure}[hbt]
\begin{minipage}{\linewidth}
\includegraphics[width=0.4\linewidth
               ]{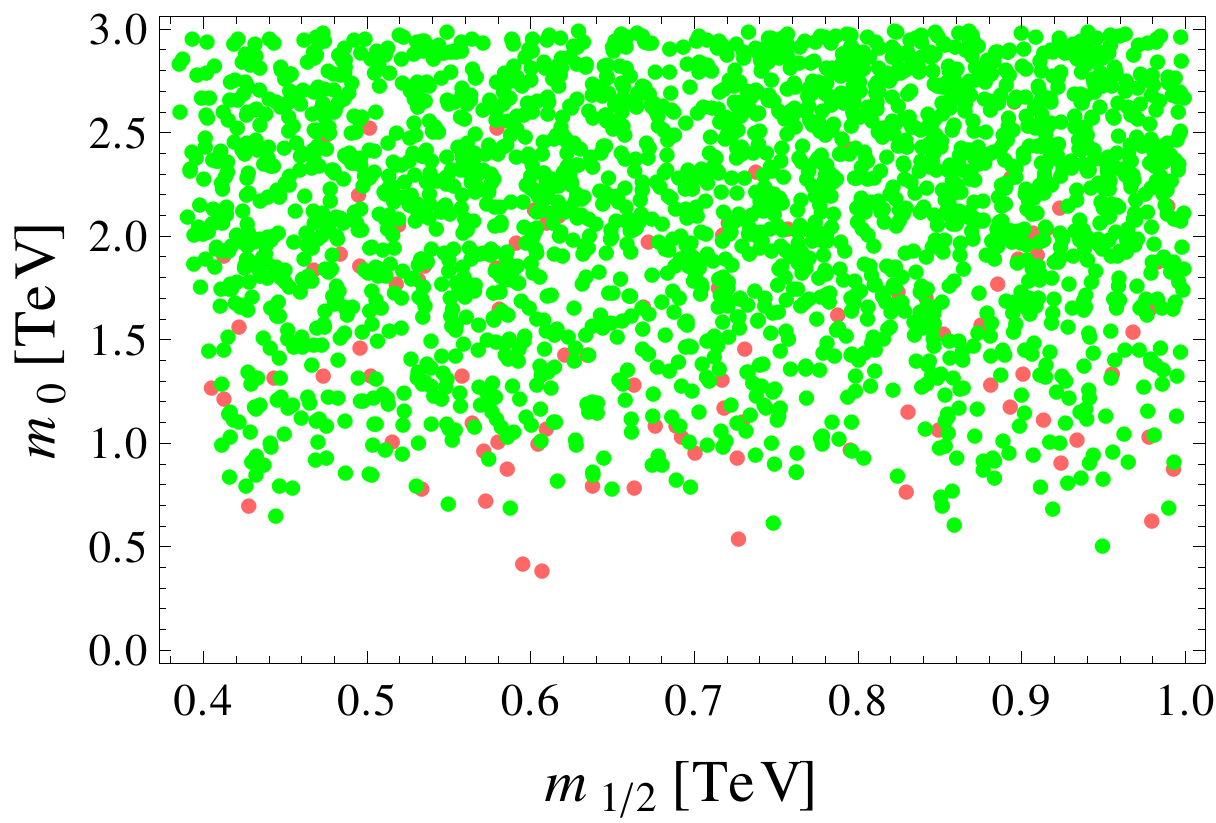}  
\hfill
\includegraphics[width=0.4\linewidth
               ]{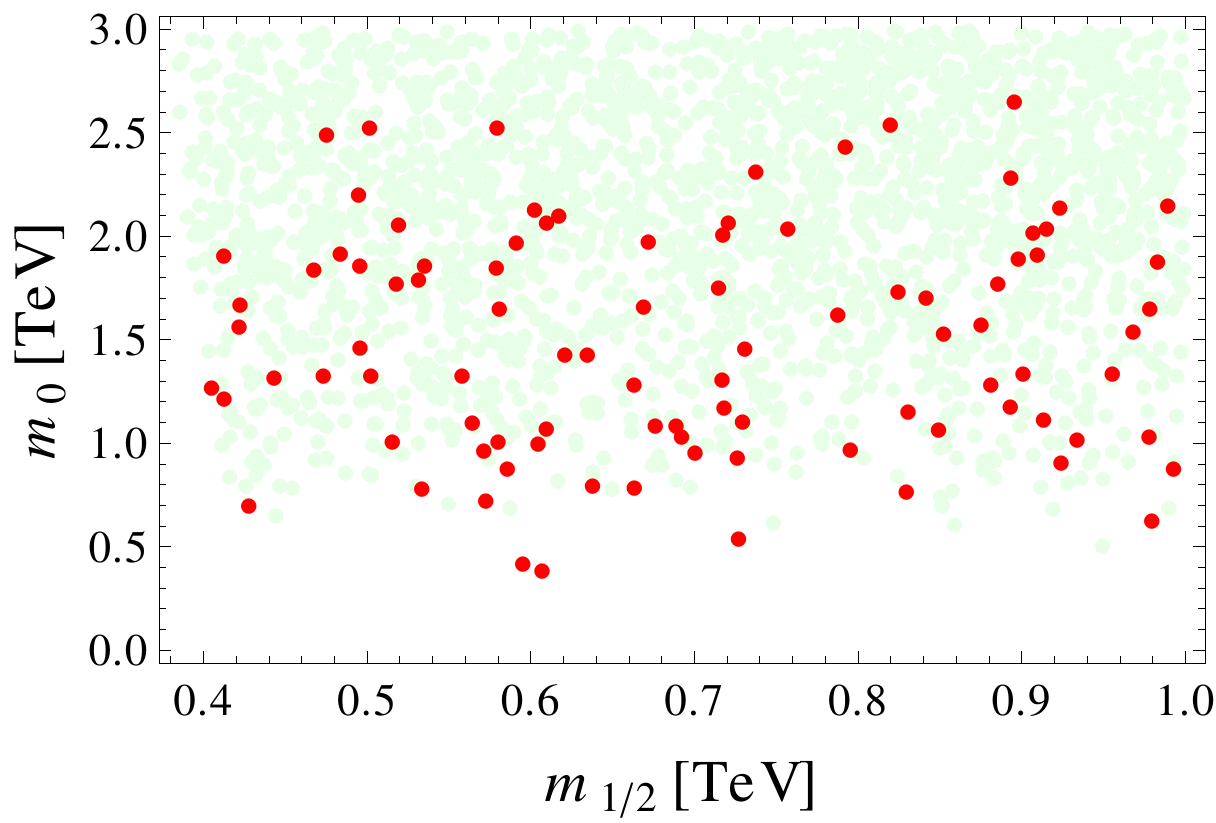}  \\
\vspace{4mm}
\includegraphics[width=0.4\linewidth
               ]{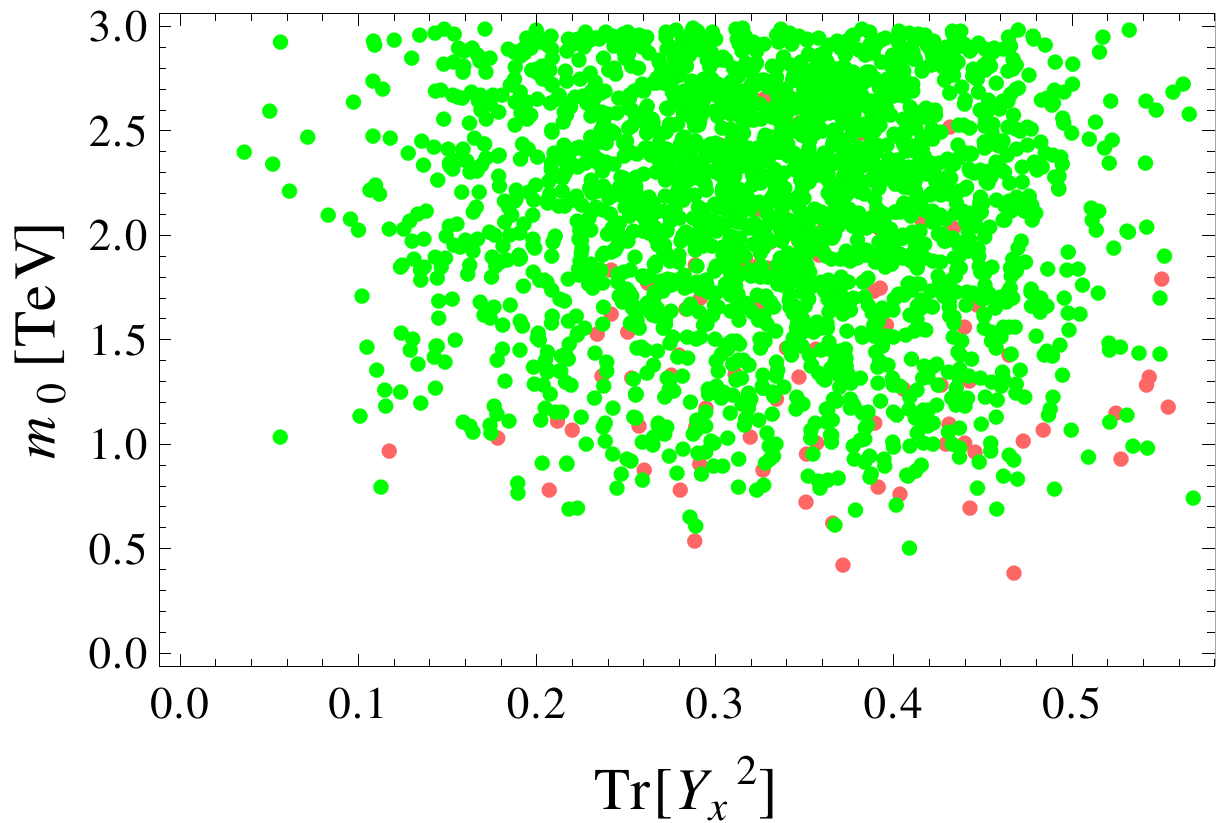}  
\hfill
\includegraphics[width=0.4\linewidth
               ]{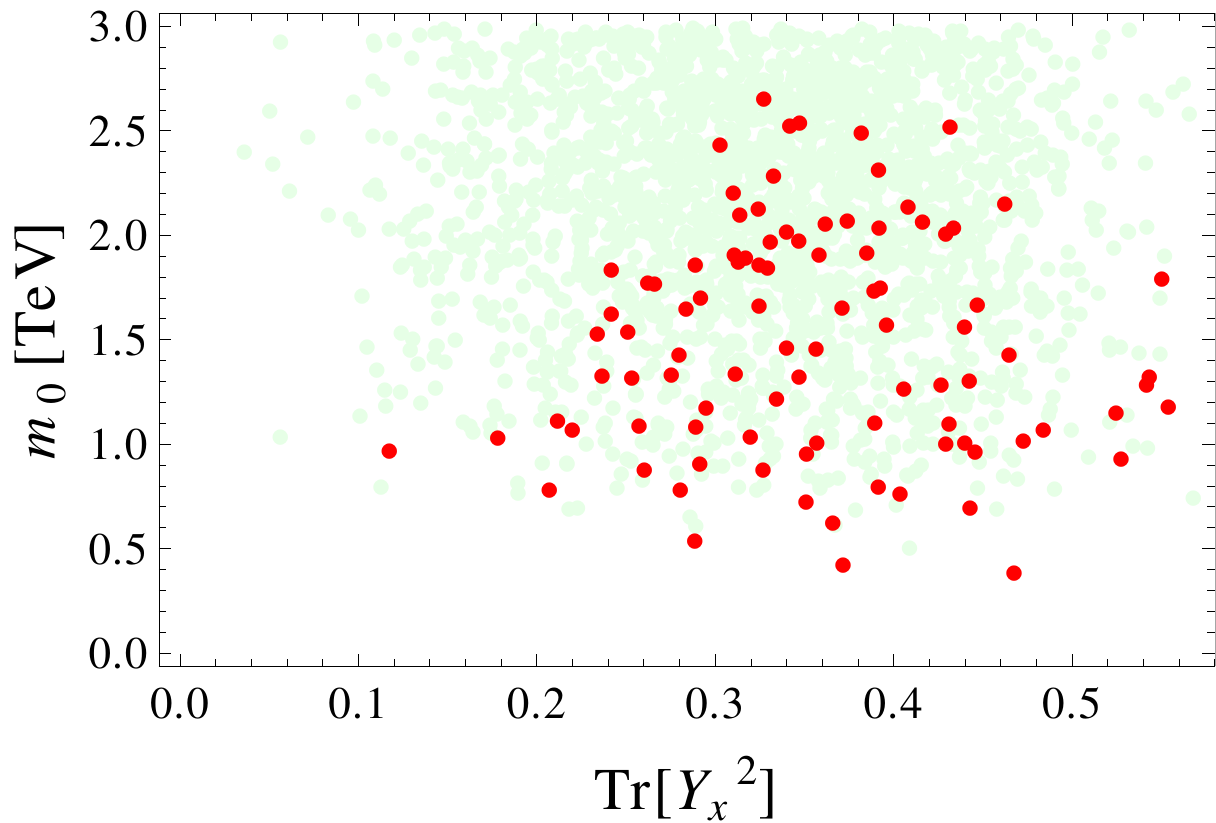}  \\
\vspace{4mm}
\includegraphics[width=0.4\linewidth
               ]{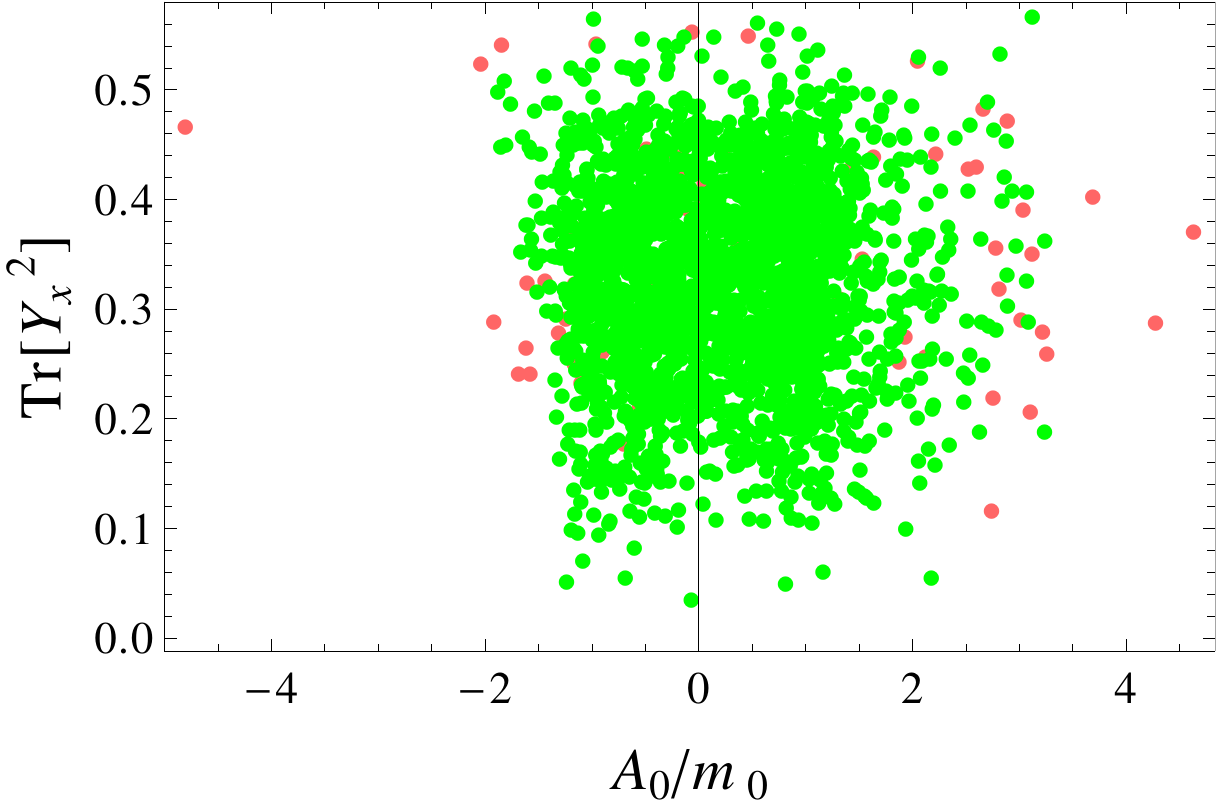}  
\hfill
\includegraphics[width=0.4\linewidth
               ]{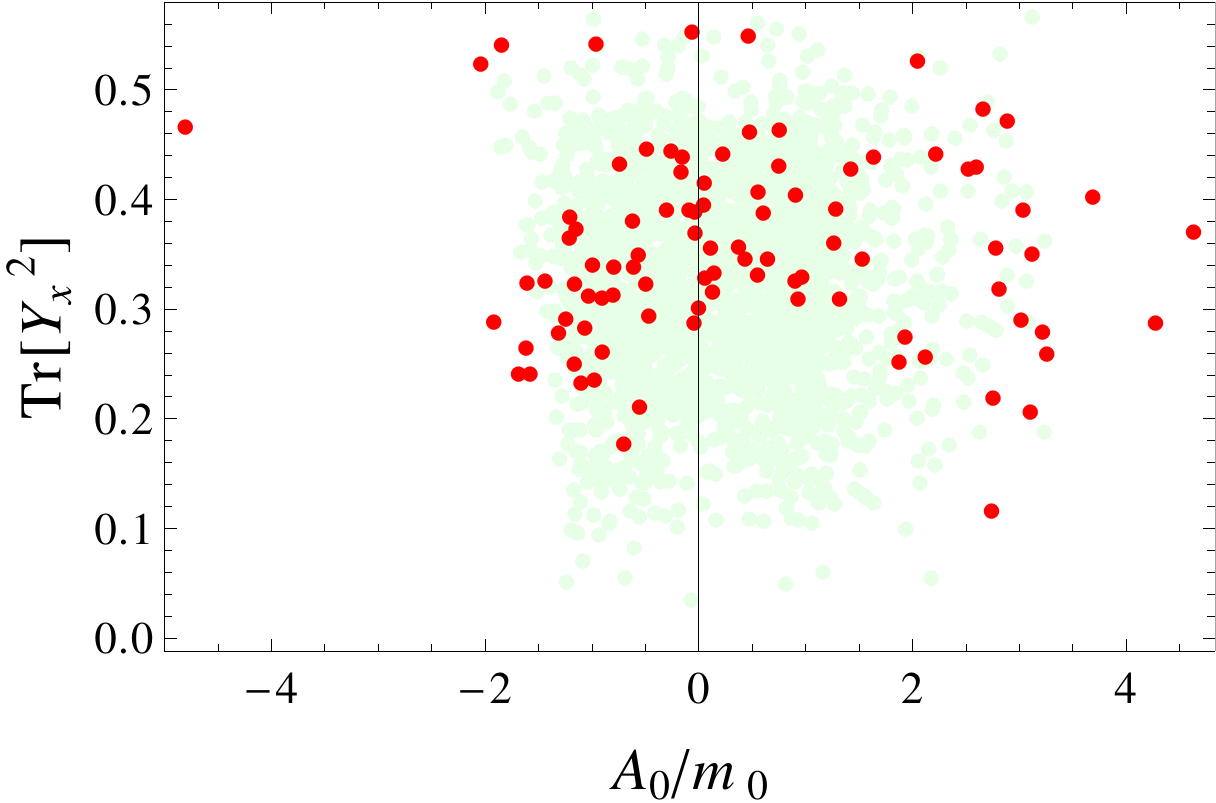}  \\
\vspace{4mm}
\includegraphics[width=0.4\linewidth
               ]{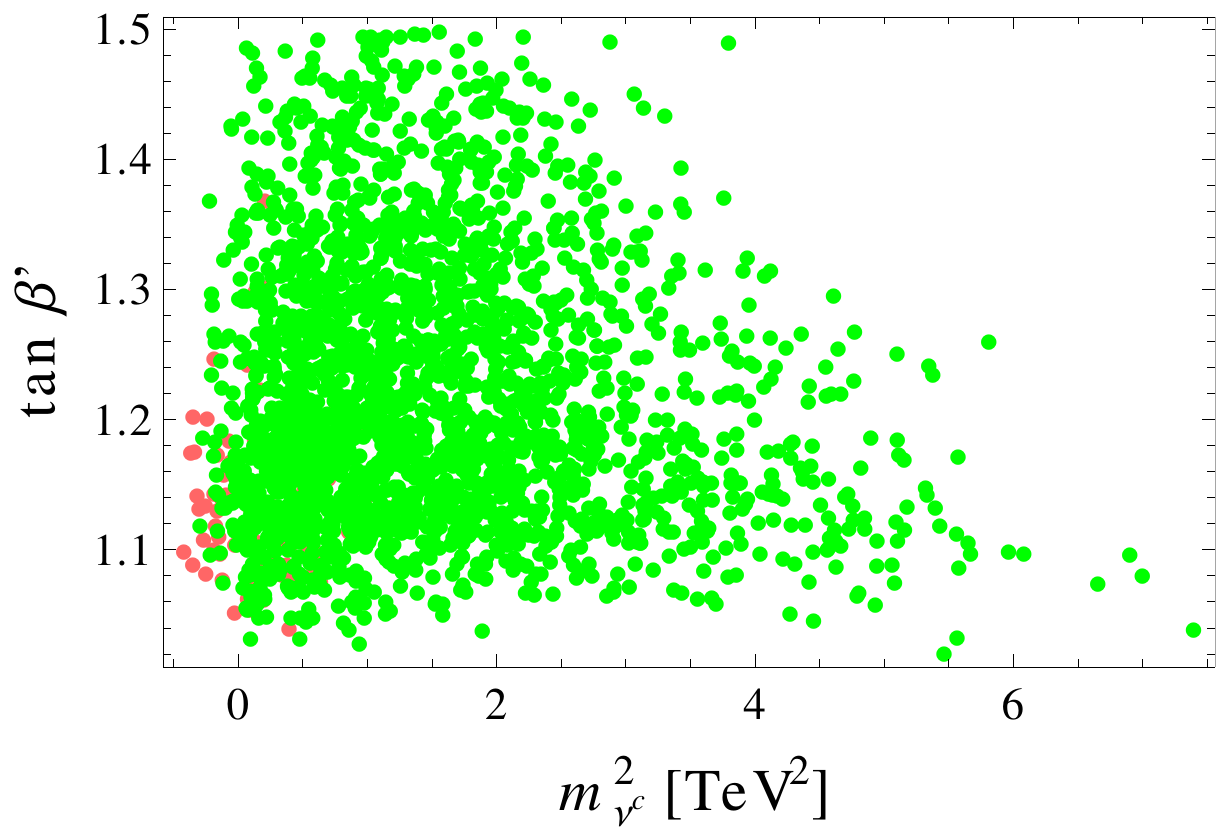}  
\hfill
\includegraphics[width=0.4\linewidth
               ]{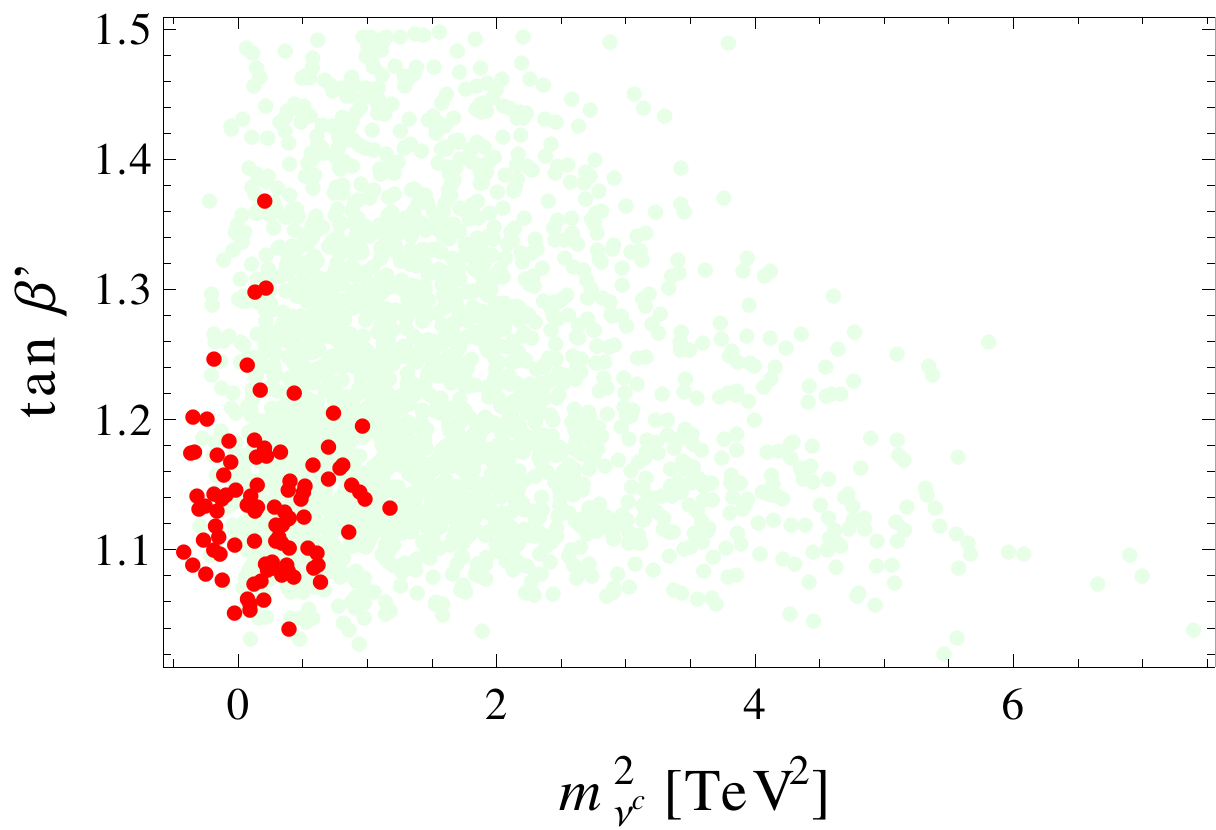}
  \end{minipage}
\caption{Projections into various parameter planes of the 2330 democratic scan
 parameter points, categorized by the nature of their global minima (see
 \TAB~\ref{tab:categorization}) at tree level.
 The scheme is the same as in
 \FIG~\ref{fig:hier_tree_comparisons}.}
\label{fig:demo_tree_comparisons}
\end{figure}

\begin{figure}[hbt]
\begin{minipage}{\linewidth}
\includegraphics[width=0.4\linewidth
               ]{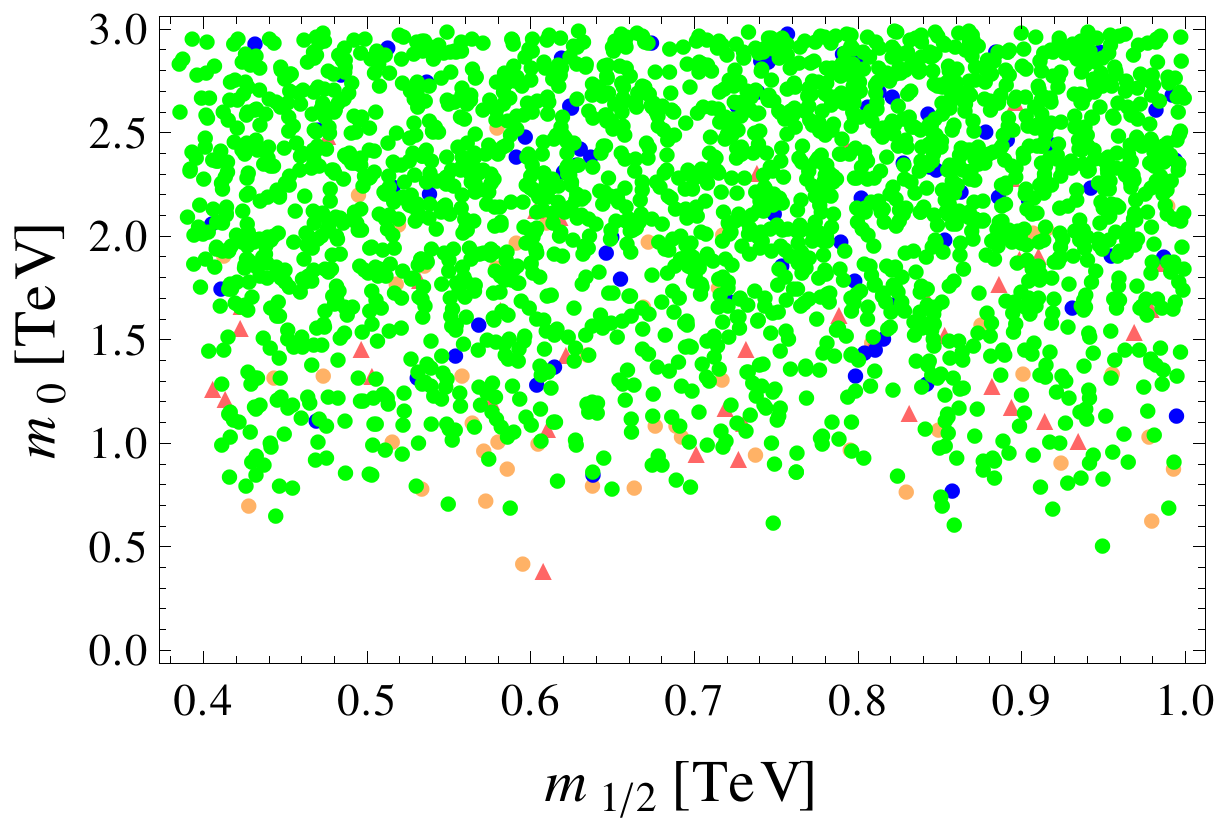}  
\hfill
\includegraphics[width=0.4\linewidth
               ]{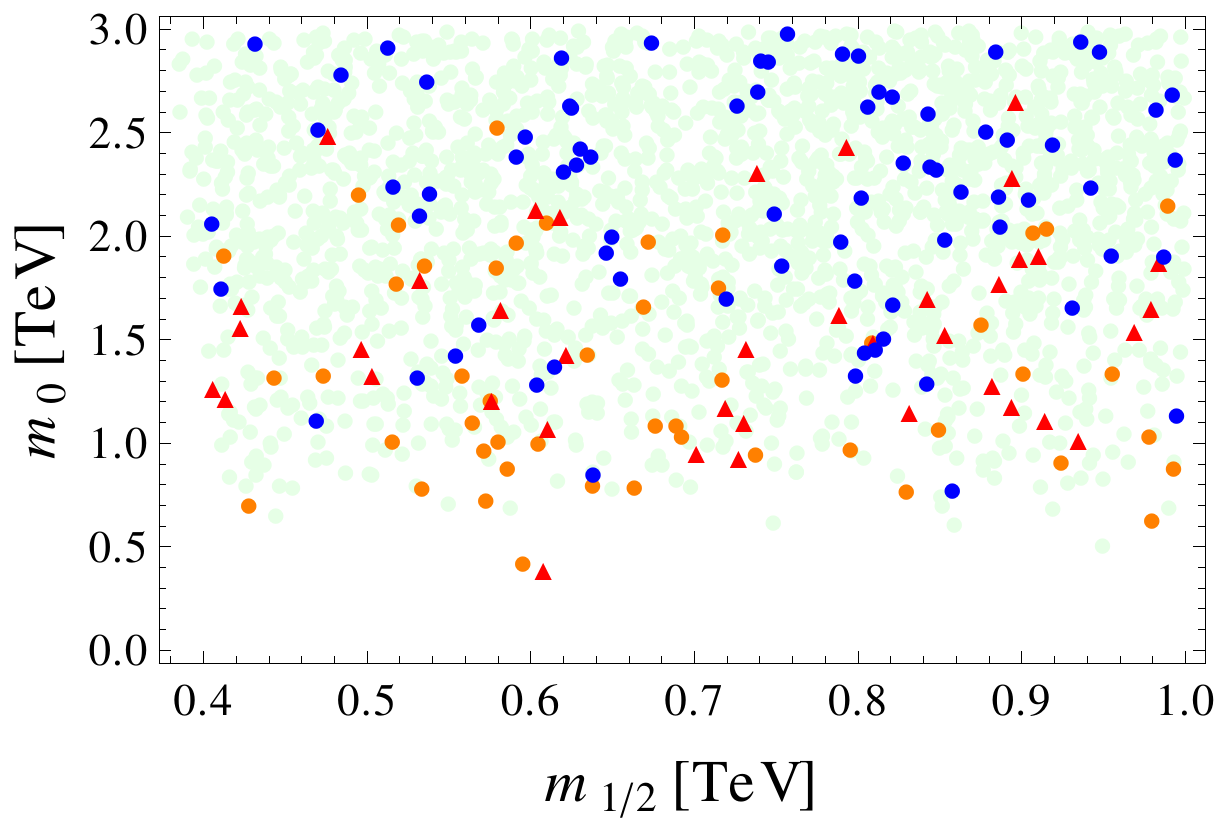}  \\
\vspace{4mm}
\includegraphics[width=0.4\linewidth
               ]{hier_loop_m0_Yx_input_stays.pdf}  
\hfill
\includegraphics[width=0.4\linewidth
               ]{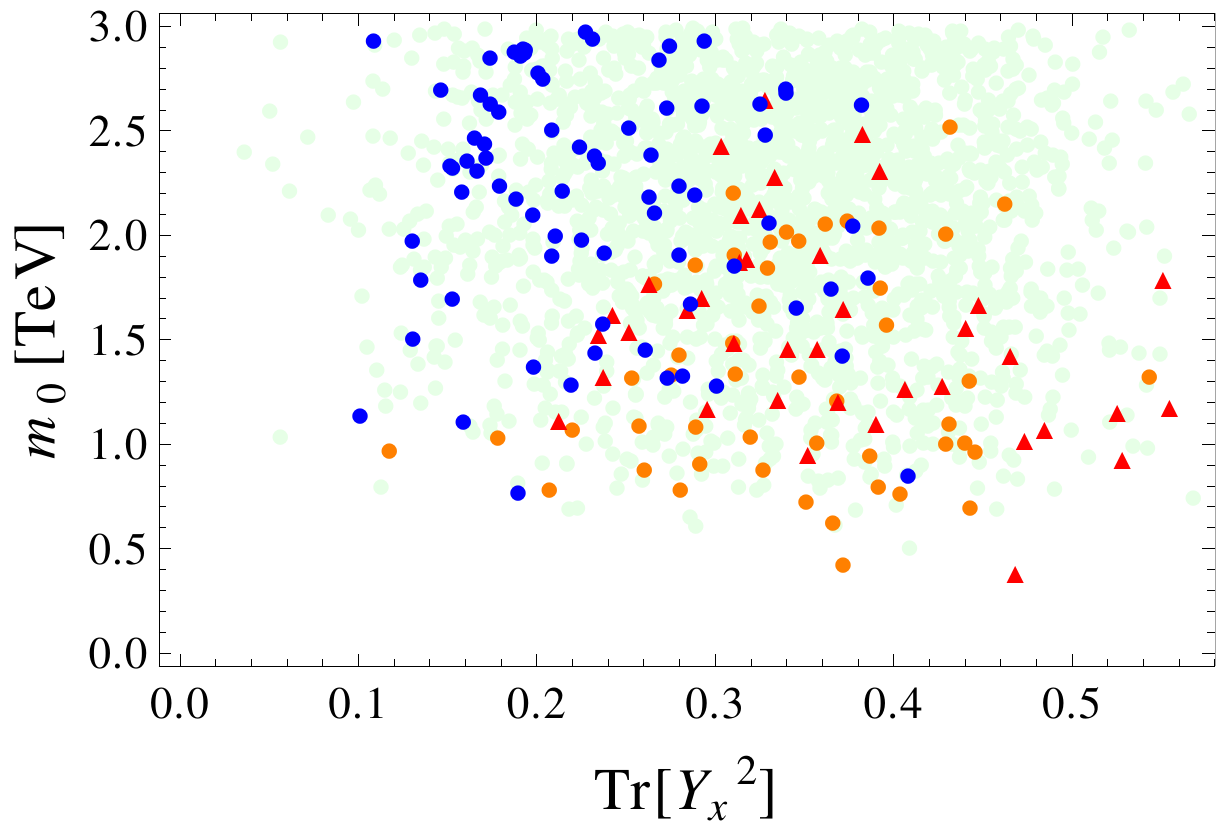}  \\
\vspace{4mm}
\includegraphics[width=0.4\linewidth
               ]{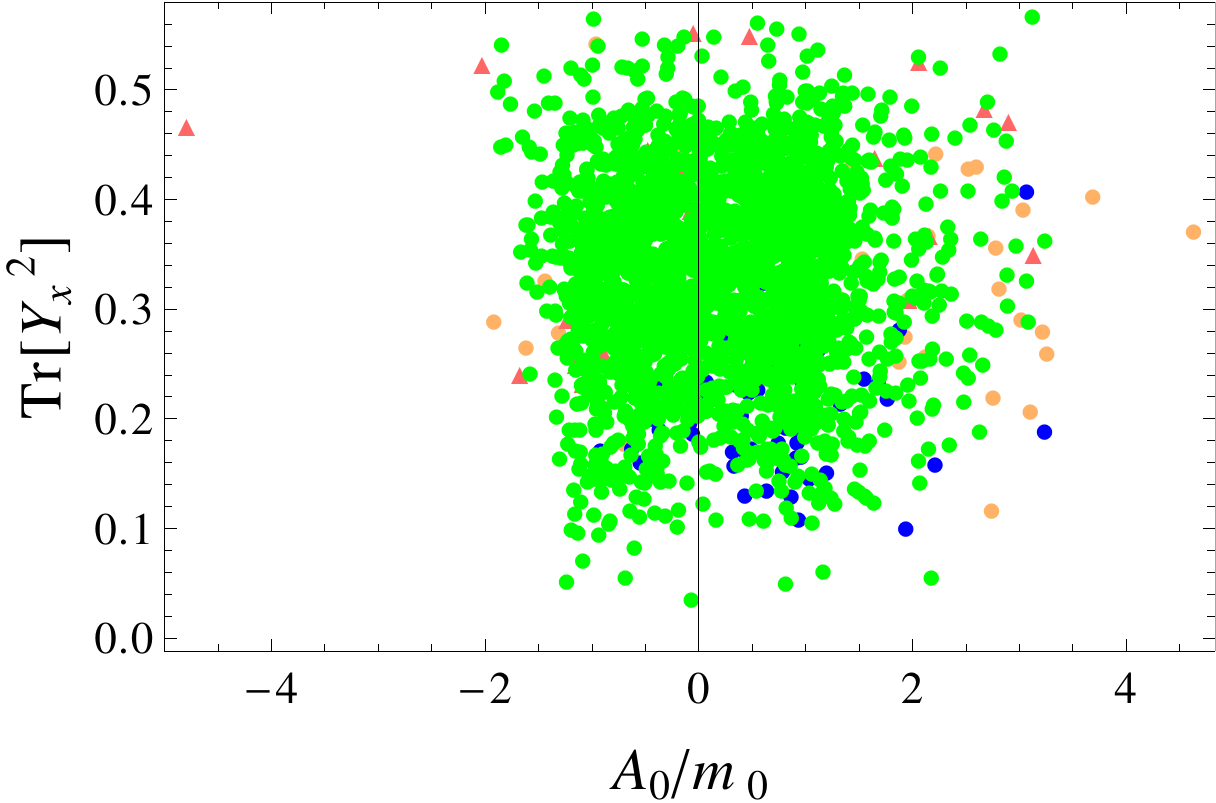}  
\hfill
\includegraphics[width=0.4\linewidth
               ]{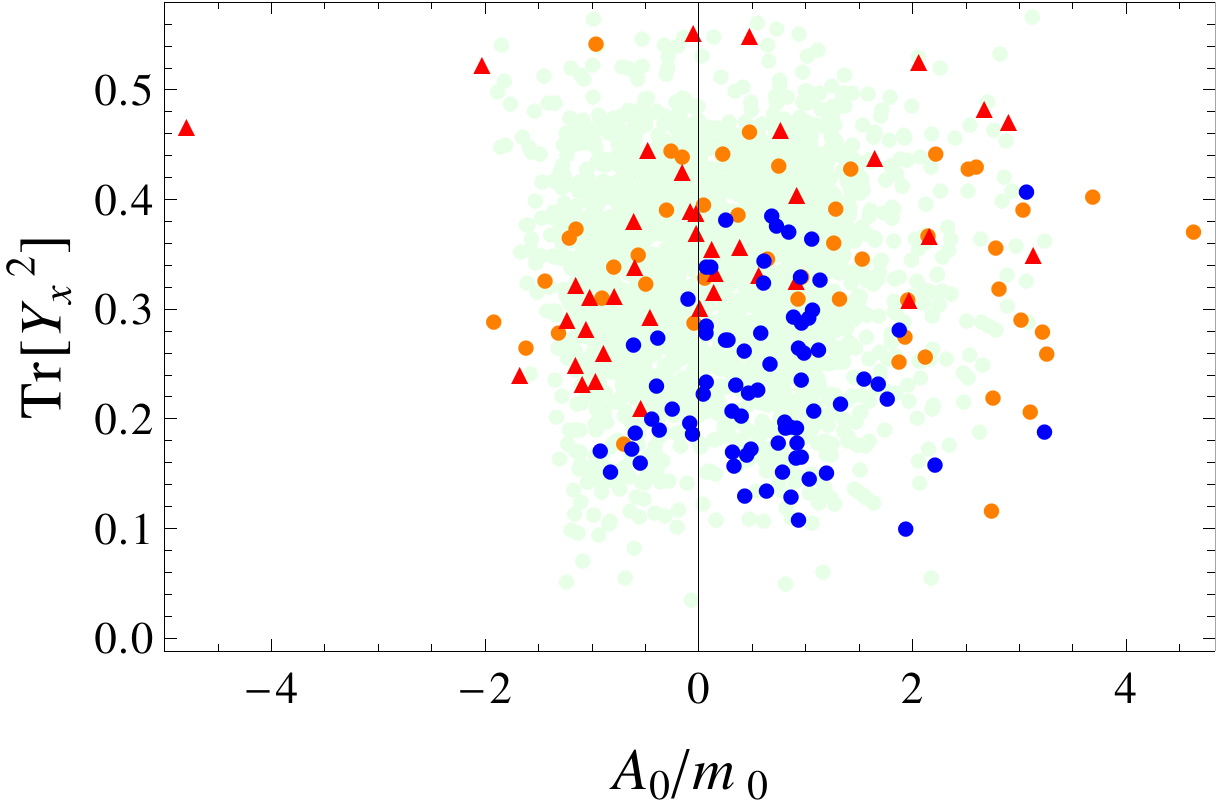}  \\
\vspace{4mm}
\includegraphics[width=0.4\linewidth
               ]{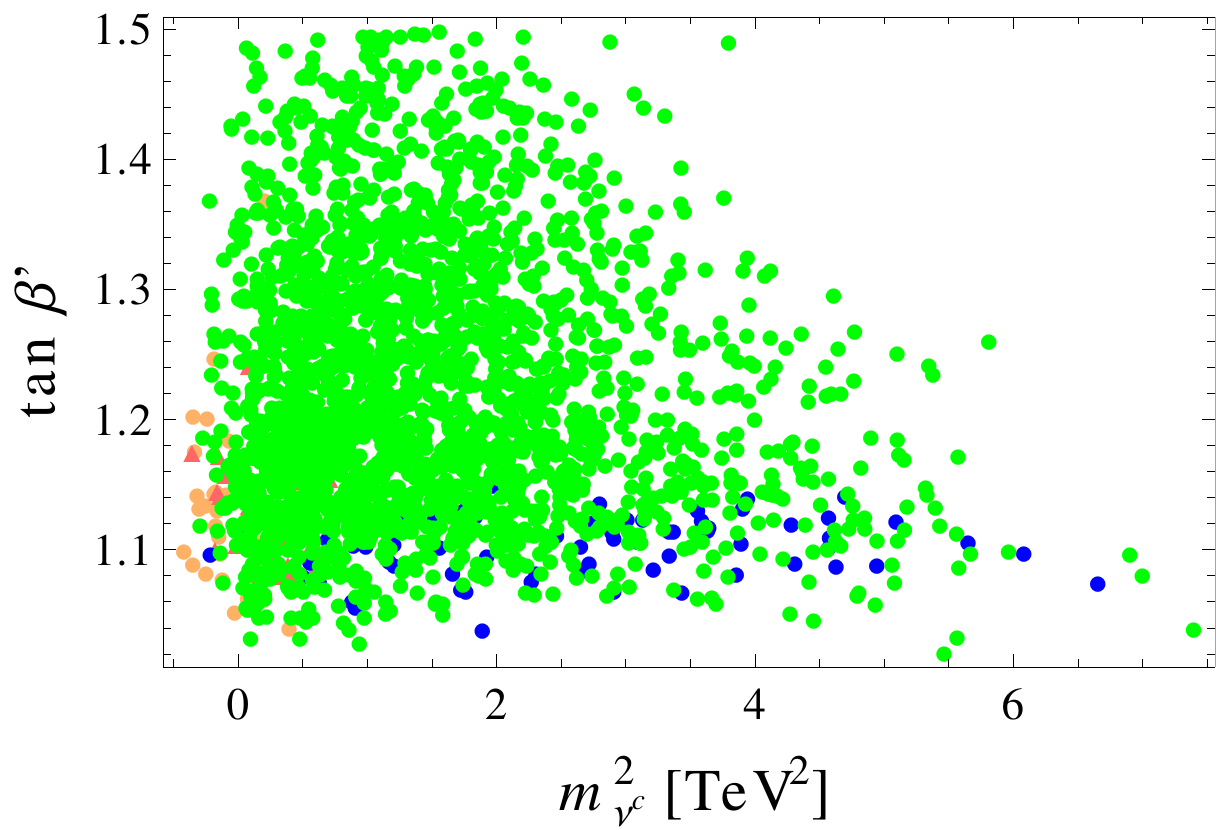}  
\hfill
\includegraphics[width=0.4\linewidth
               ]{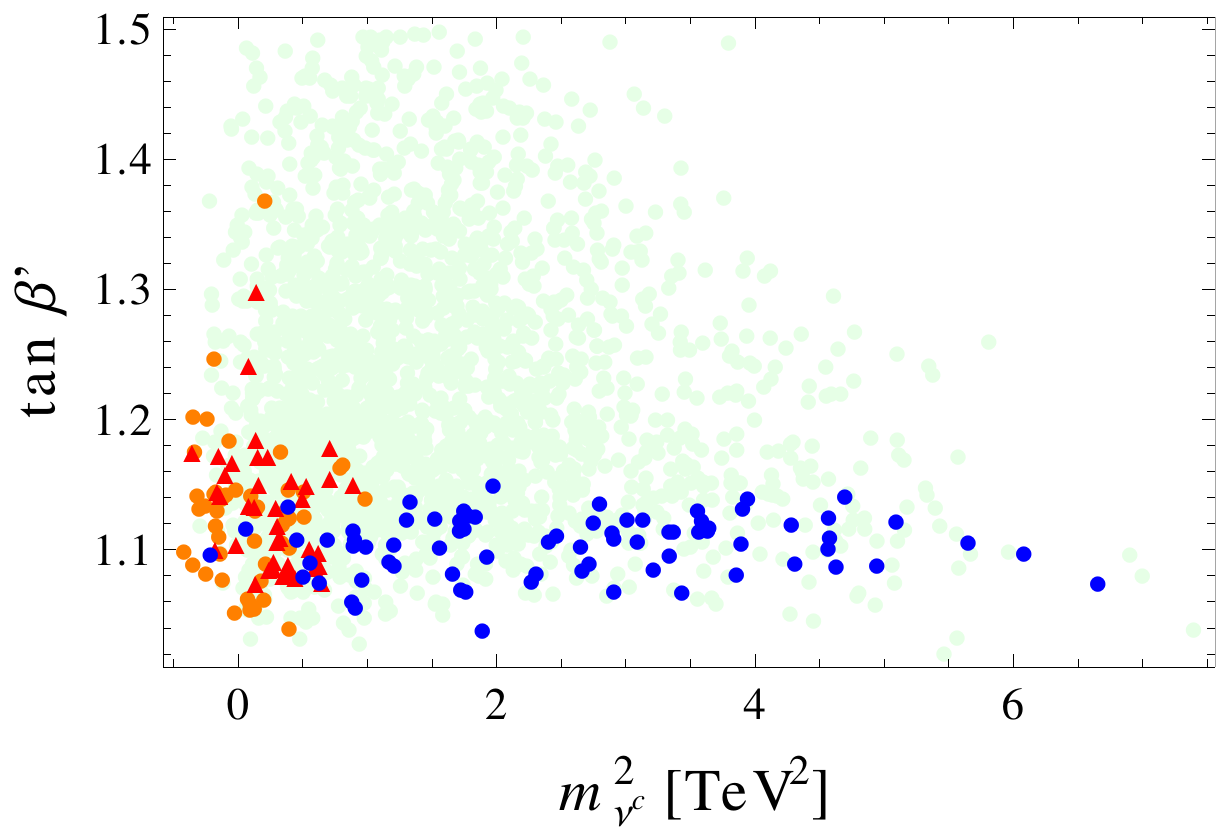}
  \end{minipage}
\caption{Projections into various parameter planes of the 2330 democratic scan
 parameter points, categorized by the nature of their global minima (see
 \TAB~\ref{tab:categorization}) at one-loop level. The scheme is the same as in
 \FIG~\ref{fig:hier_loop_comparisons}.}
\label{fig:demo_loop_comparisons}
\end{figure}

As can be seen in
 \FIGS~\ref{fig:hier_tree_comparisons}-\ref{fig:demo_loop_comparisons}, there
 are \RPCg points all over the parameter space. Based purely on a tree-level
 analysis, one might conclude that there are clear regions where the BLSSM has
 a stable, \RPCg vacuum with the correct broken and unbroken gauge groups. These
 are, as one might expect, in regions where the \rsnu-bilepton Yukawa coupling
 $Y_{x}$ is not so large, and the trilinear soft SUSY-breaking parameter $A_{0}$
 is not large compared to the soft SUSY-breaking scalar mass parameter $m_{0}$,
 as, intuitively, large $Y_{x}$ and $A_{0}$ can lead to large negative
 contributions to the potential energy for large values of $v_{R}$ and
 $v_{\eta/\bar{\eta}}$, as well as reducing the effective \rsnu mass term.
 Counter to this, a higher $m_{0}$ leads to a higher $m_{\nu}^{2}$ which, if
 positive, tends to penalize high $v_{R}$ values. This pattern is seen in both
 the hierarchical and democratic sets.
However, these conclusions run into trouble when loop corrections are taken
 into account: the regions where \RP appears to be safe at tree level apparently
 have very finely-tuned breaking of \SUL and \UBL which often does not survive
 loop corrections. Hence, while parameter points often preserved \RP, there are
 points all over the region where the global vacuum is not the
 phenomenologically-acceptable vacuum given as input, and this other vacuum is
 not trivial to find. It turns out that besides the known large contributions
 of third generation sfermions and fermions, the additional new particles
 of the $B-L$ sector also play an important role. The main reason for this is that
 the experimental
 bounds require $m_{Z'}$ to be in the multi-TeV range, implying that
 $v_\eta$ and
 $v_{\bar\eta}$ are also in this range. For $\tan\beta'\ne 1$ these \vevs give
 SUSY-breaking
 D-term contributions to the masses and, as they are much larger than the MSSM
 sector, this results in the observed importance of the corresponding loop
 contributions.
 These contributions are also responsible for the observed restoration of \UBL
 at the one-loop level.
  Moreover, as discussed in \cite{FileviezPerez:2010ek,OLeary:2011yq}
 at least one entry of $Y_x$ has to be large to achieve the breaking of \UBL.
Here we have found that one-loop contributions due to the additional $B-L$
 sector are more important in the hierarchical case, driving a larger set of
 points from the global tree-level \RPCg minimum to the global \RPVg minimum at
 one-loop level. The main reason for this is that now a single
entry has to play the role of the trace and thus is correspondingly larger
 than the average of the democratic scan.

\subsection{Comparison with previous work}
\label{sec:fileviezspinnercomparison}

Previous studies on how much of the \BLSSM parameter space conserves \RP
 \cite{FileviezPerez:2010ek}
 considered the positivity of all the soft SUSY-breaking mass-squareds of the
 \rsnus to be the necessary and sufficient condition. However, as can be seen in
 \FIGS~\ref{fig:hier_loop_comparisons} and \ref{fig:demo_loop_comparisons}, this
 is {\em neither} necessary {\em nor} sufficient. There are, of course, obvious
 trends, and parameter points with very low or negative $\mRSq$ are
 as likely to break \RP as conserve it, but there are clearly parameter points
 which conserve \RP despite having a negative $\mRSq$.
 This can be  understood by the fact that one has to consider the eigenvalues of 
 the mass matrix which can be all positive because of contributions of the
 F-terms, D-terms and other soft SUSY-breaking terms for non-zero bilepton \vevs
 despite negative $\mRSq$ entries.
 For instance, the $( {\tilde{\nu}}^{c}_{3}, {\tilde{\nu}}^{c}_{3} )$
 element of the tree-level scalar mass-squared matrix is given by
\begin{eqnarray}
m^{2}_{( {\tilde{\nu}}^{c}_{3}, {\tilde{\nu}}^{c}_{3} )}
 & = &
 m_{\nu^c}^{2} + \frac{1}{2} \left( v_{L}^{2} + v_{u}^{2} \right) |Y_\nu|^2
 - \sqrt{2} v_{\bar{\eta}} \Re\left( Y_x {\mu}^{\prime\ast} \right)
 + \sqrt{2} v_{\eta} \Re\left( T_x \right)
  +  \left( 2 v_{\eta}^{2} + 3 v_{R}^{2} \right) |Y_x|^2 \nonumber \\
 && + \frac{1}{8} \Big(
  \gmix \gBL{} \left(  v_{u}^{2} - v_{d}^{2} - v_{L}^{2} \right)
 + \gBL{2} \left( 2 \left(  v_{\bar{\eta}}^{2} - v_{\eta}^{2} \right)
 + 3 v_{R}^{2} - v_{L}^{2} \right) \Big).
\end{eqnarray}

Parameter points that have all positive $\mRSq$ yet still have \RPVg
 global minima are less surprising, as the presence of trilinear terms obviously
 could dominate for reasonably large values of sneutrino and bilepton \vevs.

Furthermore, \REF~\cite{FileviezPerez:2010ek} concluded that a hierarchical
 $Y_{x}$ would always break \RP, because of the way the couplings enter the RGEs
 and drive the soft SUSY-breaking sneutrino mass-squared parameters negative
 before the bilepton mass-squared parameters, unless all three generations
 contribute substantially. The large fraction of ``RPC'' points in our
 hierarchical scan refutes this claim. Such points are harder to find (as
 evidenced by the smaller size of the hierarchical scan to the democratic scan),
 and in some sense rather finely-tuned, as evidenced by the large fraction of
 ``unbroken'' points where loop corrections to the differences in potential are
 as large as the differences leading to the breaking of the symmetries at tree
 level. Still, almost half of the hierarchical points conserve \RP even at the
 one-loop level.

One should note, however, the different emphasis: we generated points that
 were na{\"{\i}}vely \RPCg, and explored whether they had global minima
 elsewhere, which would break \RP or not break \UBL or \SUL,
 scanning over both the soft SUSY-breaking parameters {\em and} the
 additional Yukawa couplings;
 \REF~\cite{FileviezPerez:2010ek} explored various GUT-scale parameter
 configurations in the Yukawa sector for two points in the soft
 SUSY-breaking parameters to see whether RGE running would lead to a negative
 $m_{\nu}^{2}$. Since we started with a set of parameters engineered to have an
 \RPCg local minimum, perhaps it is not so surprising to find that we did not
 find as strong a tendency for \RPVn. However, we also feel that the criterion
 used in \REF~\cite{FileviezPerez:2010ek} is not in perfect correspondance with
 whether or not the parameter point breaks or preserves \RP at its global
 minimum at the SUSY scale. 
Moreover, we have shown that loop effects do play an important role and
should not be neglected.

Finally, we examined whether the ``RPC'' minima of the ``RPV'' points were
 long-lived with respect to the age of the Universe. Unfortunately there seem to
 be no clear correlations, as can be seen in
 \FIGS~\ref{fig:hier_tree_comparisons} to \ref{fig:demo_loop_comparisons};
 we note that the SUSY-scale sign of $m_{\nu}^{2}$ appears to have little
 bearing even on the tunneling time from the ``RPC'' vacuum to the ``RPV''
 vacuum.

\section{Conclusion}
\label{sec:conclusion}

We have investigated the stability of \RP in a phenomenologically interesting
 region of the parameter space of the \BLSSM by considering all the extrema of
 the tree-level potential and using the full one-loop evaluation of the
 potential at its extrema near those of the tree-level potential. We found that
 a significant fraction of points that were chosen to conserve \RP while having
 the correct mass for the weak vector bosons actually have global minima which
 violate \RP through non-zero sneutrino \vevs. However, we also found that more
 parameter points conserve \RP than violate it.

We also found that both the conservation and violation of \RP were possible all
 over the parameter space, in contradiction to previous studies, and while
 trends in the parameters leading either case are visible, it is difficult to
 identify in advance what values of which parameters will lead definitely to the
 conservation or violation of \RP.

\section*{Acknowledgments}
This work has been supported by the DFG, project No. PO-1337/2-1, and
partly by the Helmholtz alliance `Physics at the Terascale'.
BO'L has been supported by DFG research training group GRK 1147.

\appendix

\section{The one-loop effective potential}
\label{app:effpot}
The effective potential at one-loop is calculated by using \EQ~(\ref{eq:EffPot})
 and reads in its full form:
\begin{align}
\nonumber 16 \pi^2 &(V^1 - V^0) = -3 \left(\sum_{i=1}^3
\left(\frac{m^4_{d_i}}{4}\left[ln \left(\frac{m^2_{d_i}}{Q^2}\right) -
\frac{3}{2}\right]\right) +
 \sum_{i=1}^3 \left(\frac{m^4_{u_i}}{4}\left[ln \left(
\frac{m^2_{u_i}}{Q^2}\right) - \frac{3}{2}\right]\right)\right)  \\
\nonumber & + \frac{3}{2} \left(\sum_{i=1}^6
\left(\frac{m^4_{\tilde{d}_i}}{4}\left[ln \left(
\frac{m^2_{\tilde{d}_i}}{Q^2}\right) - \frac{3}{2}\right]\right) +
 \sum_{i=1}^6 \left(\frac{m^4_{\tilde{u}_i}}{4}\left[ln \left(
\frac{m^2_{\tilde{u}_i}}{Q^2}\right) - \frac{3}{2}\right]\right) \right)  \\
\nonumber & + \frac{3}{4}\left(\frac{m^4_W}{2}\left[ln \left(
\frac{m^2_W}{Q^2}\right) + \frac{3}{2}\right] + \frac{m^4_Z}{4}\left[ln \left(
\frac{m^2_Z}{Q^2}\right) - 
\frac{3}{2}\right]+ \frac{m^4_{Z'}}{4}\left[ln \left(
\frac{m^2_{Z'}}{Q^2}\right) - \frac{3}{2}\right]\right)  \\
\nonumber & - \sum_{i=1}^5 \left(\frac{m^4_{\tilde{\chi}^\pm_i}}{4}\left[ln
\left( \frac{m^2_{\tilde{\chi}^\pm_i}}{Q^2}\right) - \frac{3}{2}\right]\right)
- 
 \frac{1}{2} \sum_{i=1}^{13} \left(\frac{m^4_{\tilde{\chi}^0_i}}{4}\left[ln
\left( \frac{m^2_{\tilde{\chi}^0_i}}{Q^2}\right) - \frac{3}{2}\right]\right)  \\
\nonumber  &+  \frac{1}{4} \left(\sum_{i=1}^{10}
\left(\frac{m^4_{h_i}}{4}\left[ln \left( \frac{m^2_{h_i}}{Q^2}\right) -
\frac{3}{2}\right]\right) +
 \sum_{i=1}^{10} \left(\frac{m^4_{A^0_i}}{4}\left[ln \left(
\frac{m^2_{A^0_i}}{Q^2}\right) - \frac{3}{2}\right]\right) \right)  \\&+
 \frac{1}{2} \sum_{i=1}^8 \left(\frac{m^4_{\tilde{e}_i}}{4}\left[ln \left(
\frac{m^2_{\tilde{e}_i}}{Q^2}\right) - \frac{3}{2}\right]\right).
\end{align}

The first two lines are the contributions from quarks and squarks which are the
same as for the MSSM. In the third line the loops including the three gauge
bosons in the given model are counted. The fourth line contains the contributions
due to the charged lepton-chargino as well as neutrino-neutralino mixed states.
The fifth and sixth lines show the contributions from charged slepton-charged
Higgsino states as well as sneutrino-Higgs states. We have included formally in
 the sums the would-be Goldstone-bosons to account for the fact that potentially
 the gauge group is broken only partially. This does not result in a
 double-counting because, in the Landau gauge, would-be Goldstone-bosons are
 massless and thus give a zero contribution to the one-loop effective potential.
 The tree-level mass matrices where \RPVn induces a mixing between SM and SUSY
 particles are listed in appendix~\ref{app:massmatrices}. For completeness we
 note that in addition the sneutrino \vevs give contributions to the mass
 matrices of vector bosons and squarks. 

\section{Mass matrices}
\allowdisplaybreaks
\label{app:massmatrices}
Here we give the tree-level masses suppressing the generation indices of
 (s)neutrinos and (s)leptons.
\begin{itemize}
\item {\bf Neutrino-Neutralino states}\\
The mass matrix $m^2_{\tilde{\chi}^0}$ is given, in the basis
 $$ \left(\lambda_{\tilde{B}}, \tilde{W}^0, \tilde{H}_d^0, \tilde{H}_u^0, \nu_L,
\nu_R^*, \tilde{\eta}, \tilde{\bar{\eta}}, \lambda_{\tilde{B'}}\right), $$
by

\begin{equation} 
\scriptscriptstyle{\left( 
\begin{array}{ccccccccc}
M_1 &0 &-\frac{1}{2} g_1 v_d  &\frac{1}{2} g_1 v_u  &-\frac{1}{2} g_1 v_L  &0 &0
&0 & {M}_{B B'} \\ 
0 &M_2 &\frac{1}{2} g_2 v_d  &-\frac{1}{2} g_2 v_u  &\frac{1}{2} g_2 v_L  &0 &0
&0 &0\\ 
-\frac{1}{2} g_1 v_d  &\frac{1}{2} g_2 v_d  &0 &- \mu  &0 &0 &0 &0 &-\frac{1}{2}
\bar{g} v_d \\ 
\frac{1}{2} g_1 v_u  &-\frac{1}{2} g_2 v_u  &- \mu  &0 &\frac{1}{\sqrt{2}} v_R
Y_\nu  &\frac{1}{\sqrt{2}} v_L Y_\nu  &0 &0 &\frac{1}{2} \bar{g} v_u \\ 
-\frac{1}{2} g_1 v_L  &\frac{1}{2} g_2 v_L  &0 &\frac{1}{\sqrt{2}} v_R Y_\nu  &0
&\frac{1}{\sqrt{2}} v_u Y_\nu  &0 &0 &-\frac{1}{2} (\bar{g} + g_{B})v_L\\ 
0 &0 &0 &\frac{1}{\sqrt{2}} v_L Y_\nu  &\frac{1}{\sqrt{2}} v_u Y_\nu  &\sqrt{2} v_{\eta} Y_x
&\sqrt{2} v_R Y_x &0 &\frac{1}{2} g_{B} v_R \\ 
0 &0 &0 &0 &0 &\sqrt{2} v_R Y_x &0 &- {\mu'}  &- g_{B} v_{\eta} \\ 
0 &0 &0 &0 &0 &0 &- {\mu'}  &0 &g_{B} v_{\bar{\eta}} \\ 
{M}_{B B'}  &0 &-\frac{1}{2} \bar{g} v_d  &\frac{1}{2} \bar{g} v_u 
&-\frac{1}{2} (\bar{g} + g_{B})v_L &\frac{1}{2} g_{B} v_R  &- g_{B} v_{\eta}  &g_{B} v_{\bar{\eta}} 
&{M}_{B'}\end{array} 
\right)}.
\end{equation}

\item {\bf Charged lepton-Charginos}\\
The mass matrix is given, in the basis
 $$ \left(e_L, \tilde{W}^-,\tilde{H}_d^-\right); \left(e_R^*, \tilde{W}^+, \tilde{H}_u^+\right), $$   
by
\begin{equation} 
m^2_{\tilde{\chi}^-} = \left( 
\begin{array}{ccc}
\frac{1}{\sqrt{2}} v_d Y_{e }  &\frac{1}{\sqrt{2}} g_2 v_L  &-
\frac{1}{\sqrt{2}} v_R Y_\nu \\ 
0 &M_2 &\frac{1}{\sqrt{2}} g_2 v_u \\ 
- \frac{1}{\sqrt{2}} v_L Y_{e }  &\frac{1}{\sqrt{2}} g_2 v_d 
&\mu\end{array} 
\right) .
\end{equation}

\item {\bf CP even sneutrino-Higgs}\\
In the basis $$ \left(\sigma_{d}, \sigma_{u}, \sigma_L, \sigma_R, \sigma_\eta,
\sigma_{\bar{\eta}}\right) ,$$ the entries of the mass matrix read:
\begin{align} 
m_{\sigma_{d}\sigma_{d}} &= \frac{1}{8} \Big(\bar{g} g_{BL} \Big(2 v_{\eta}^{2} -2 v_{\bar{\eta}}^{2} - v_{R}^{2}  + v_{L}^{2}\Big) + \Big(g_{1}^{2} + \bar{g}^{2} + g_{2}^{2}\Big)\Big(3 v_{d}^{2}  - v_{u}^{2}  + v_{L}^{2}\Big)\Big) + m_{H_d}^2 + \mu^{2}\\ 
m_{\sigma_{d}\sigma_{u}} &= - B_{\mu}  -\frac{1}{4} \Big(g_{1}^{2} + \bar{g}^{2} + g_{2}^{2}\Big)v_d v_u \\ 
m_{\sigma_{u}\sigma_{u}} &= m_{H_u}^2+\frac{1}{8} \Big(\Big(- \bar{g}^{2}  - g_{1}^{2}  - g_{2}^{2} \Big)\Big( v_{d}^{2} -3 v_{u}^{2} + v_{L}^{2}\Big) + \bar{g} g_{BL} \Big(2 v_{\bar{\eta}}^{2}  -2 v_{\eta}^{2}  - v_{L}^{2}  + v_{R}^{2}\Big)\Big) \nonumber \\
& +\frac{1}{2} \Big(v_{L}^{2} + v_{R}^{2}\Big)Y_{\nu}^{2} +\mu^{2}\\ 
m_{\sigma_{d}\sigma_L} &= \frac{1}{4} \Big(\bar{g} \Big(\bar{g} + g_{BL}\Big) + g_{1}^{2} + g_{2}^{2}\Big)v_d v_L  - \frac{1}{\sqrt{2}} v_R Y_\nu \mu \\ 
m_{\sigma_{u}\sigma_L} &= -\frac{1}{4} \Big(\bar{g} \Big(\bar{g} + g_{BL}\Big) + g_{1}^{2} + g_{2}^{2}\Big)v_L v_u  + \frac{1}{\sqrt{2}} v_R T_\nu  + Y_\nu \Big(v_L v_u Y_\nu  + v_R v_{\eta} Y_x \Big)\\ 
m_{\sigma_L\sigma_L} &= m_L^2+\frac{1}{8} \Big(\Big(g_{1}^{2} + \bar{g}^{2} + g_{2}^{2}\Big)\Big(3 v_{L}^{2}  - v_{u}^{2}  + v_{d}^{2}\Big)+g_{BL}^{2} \Big(-2 v_{\bar{\eta}}^{2}  + 2 v_{\eta}^{2}  + 3 v_{L}^{2}  - v_{R}^{2} \Big)\nonumber \\ 
 &+\bar{g} g_{BL} \Big(-2 v_{\bar{\eta}}^{2}  + 2 v_{\eta}^{2}  + 6 v_{L}^{2}  - v_{R}^{2}  - v_{u}^{2}  + v_{d}^{2}\Big)\Big)+\frac{1}{2} \Big(v_{R}^{2} + v_{u}^{2}\Big)Y_{\nu}^{2} \\ 
m_{\sigma_{d}\sigma_R} &= -\frac{1}{4} \bar{g} g_{BL} v_d v_R  - \frac{1}{\sqrt{2}} v_L Y_\nu \mu \\ 
m_{\sigma_{u}\sigma_R} &= \frac{1}{4} \bar{g} g_{BL} v_R v_u  + \frac{1}{\sqrt{2}} v_L T_\nu  + Y_\nu \Big(v_L v_{\eta} Y_x  + v_R v_u Y_\nu \Big)\\ 
m_{\sigma_L\sigma_R} &= -\frac{1}{4} g_{BL} \Big(\bar{g} + g_{BL}\Big)v_L v_R  + \frac{1}{\sqrt{2}} v_u T_\nu  + v_L v_R Y_{\nu}^{2}  + Y_\nu \Big(- \frac{1}{\sqrt{2}} v_d \mu  + v_u v_{\eta} Y_x \Big)\\ 
m_{\sigma_R\sigma_R} &= {m_{\nu^c}^2}-\frac{1}{8} g_{BL} \Big(\bar{g} \Big(- v_{u}^{2}  + v_{d}^{2} + v_{L}^{2}\Big) + g_{BL} \Big(-2 v_{\bar{\eta}}^{2}  + 2 v_{\eta}^{2}  -3 v_{R}^{2}  + v_{L}^{2}\Big)\Big)\nonumber \\ 
 &+\frac{1}{2} \Big(2 \Big(\sqrt{2} v_{\eta} T_x  + Y_x \Big(\Big(2 v_{\eta}^{2}  + 3 v_{R}^{2} \Big)Y_x  - \sqrt{2} {\mu'} v_{\bar{\eta}} \Big)\Big) + \Big(v_{L}^{2} + v_{u}^{2}\Big)Y_{\nu}^{2} \Big)\\ 
m_{\sigma_{d}\sigma_{\eta}} &= \frac{1}{2} \bar{g} g_{BL} v_d v_{\eta} \\ 
m_{\sigma_{u}\sigma_{\eta}} &= -\frac{1}{2} \bar{g} g_{BL} v_u v_{\eta}  + v_L v_R Y_x Y_\nu \\ 
m_{\sigma_L\sigma_{\eta}} &= \frac{1}{2} g_{BL} \Big(\bar{g} + g_{BL}\Big)v_L v_{\eta}  + v_R v_u Y_x Y_\nu \\ 
m_{\sigma_R\sigma_{\eta}} &= -\frac{1}{2} g_{BL}^{2} v_R v_{\eta}  + \sqrt{2} v_R T_x  + Y_x \Big(4 v_R v_{\eta} Y_x  + v_L v_u Y_\nu \Big)\\ 
m_{\sigma_{\eta}\sigma_{\eta}} &= 2 v_{R}^{2} Y_{x}^{2}  + \frac{1}{4} g_{BL} \Big(\bar{g} \Big(v_{d}^{2}- v_{u}^{2} + v_{L}^{2}\Big) + g_{BL} \Big(6 v_{\eta}^{2} -2 v_{\bar{\eta}}^{2} - v_{R}^{2}  + v_{L}^{2}\Big)\Big) + m_{\eta}^2 + {\mu'}^{2}\\ 
m_{\sigma_{d}\sigma_{\bar{\eta}}} &= -\frac{1}{2} \bar{g} g_{BL} v_d v_{\bar{\eta}} \\ 
m_{\sigma_{u}\sigma_{\bar{\eta}}} &= \frac{1}{2} \bar{g} g_{BL} v_u v_{\bar{\eta}} \\ 
m_{\sigma_L\sigma_{\bar{\eta}}} &= -\frac{1}{2} g_{BL} \Big(\bar{g} + g_{BL}\Big)v_L v_{\bar{\eta}} \\ 
m_{\sigma_R\sigma_{\bar{\eta}}} &= \frac{1}{2} g_{BL}^{2} v_R v_{\bar{\eta}}  - \sqrt{2} {\mu'} v_R Y_x \\ 
m_{\sigma_{\eta}\sigma_{\bar{\eta}}} &= - {B_{\mu'}}  - g_{BL}^{2} v_{\eta} v_{\bar{\eta}} \\ 
m_{\sigma_{\bar{\eta}}\sigma_{\bar{\eta}}} &= -\frac{1}{4} g_{BL} \Big(\bar{g} \Big(- v_{u}^{2}  + v_{d}^{2} + v_{L}^{2}\Big) + g_{BL} \Big(2 v_{\eta}^{2}  -6 v_{\bar{\eta}}^{2}  - v_{R}^{2}  + v_{L}^{2}\Big)\Big) + m_{\bar{\eta}}^2 + {\mu'}^{2}
\end{align}

\item {\bf CP odd sneutrino-Higgs}\\
In the basis \( \left(\phi_{d}, \phi_{u}, \phi_L, \phi_R, \phi_{\eta},
\phi_{\bar{\eta}}\right) \) and using Landau gauge, the nonzero entries of the
mass matrix read:

\begin{align} 
m_{\phi_{d}\phi_{d}} &= \frac{1}{8} \Big(\bar{g} g_{BL} \Big(2 \Big( v_{\eta}^{2}- v_{\bar{\eta}}^{2}\Big) - v_{R}^{2}  + v_{L}^{2}\Big) + \Big(g_{1}^{2} + \bar{g}^{2} + g_{2}^{2}\Big)\Big( v_{d}^{2}- v_{u}^{2}  + v_{L}^{2}\Big)\Big) + m_{H_d}^2 + \mu^{2}\\ 
m_{\phi_{d}\phi_{u}} &= B_{\mu}\\
m_{\phi_{u}\phi_{u}} &= m_{H_u}^2+\frac{1}{8} \Big(\Big(- \bar{g}^{2}  - g_{1}^{2}  - g_{2}^{2} \Big)\Big(- v_{u}^{2}  + v_{d}^{2} + v_{L}^{2}\Big) + \bar{g} g_{BL} \Big(2 v_{\bar{\eta}}^{2}  -2 v_{\eta}^{2}  - v_{L}^{2}  + v_{R}^{2}\Big)\Big)\nonumber \\ 
 &+\frac{1}{2} \Big(v_{L}^{2} + v_{R}^{2}\Big)Y_{\nu}^{2} +\mu^{2}\\ 
m_{\phi_{d}\phi_{L}} &= - \frac{1}{\sqrt{2}} v_R Y_\nu \mu\\
m_{\phi_{u}\phi_L} &= -\frac{1}{2} v_R \Big(2 v_{\eta} Y_x Y_\nu  + \sqrt{2} T_\nu \Big)\\ 
m_{\phi_L\phi_L} &= m_L^2+\frac{1}{8} \Big(\Big(g_{1}^{2} + \bar{g}^{2} + g_{2}^{2}\Big)\Big( v_{d}^{2}- v_{u}^{2} + v_{L}^{2}\Big)+\bar{g} g_{BL} \Big(2 \Big(v_{L}^{2} + v_{\eta}^{2} - v_{\bar{\eta}}^{2} \Big) - v_{R}^{2}  - v_{u}^{2}  + v_{d}^{2}\Big)\nonumber \\ 
 &+g_{BL}^{2} \Big(2 \Big(- v_{\bar{\eta}}^{2}  + v_{\eta}^{2}\Big) - v_{R}^{2}  + v_{L}^{2}\Big)\Big)+\frac{1}{2} \Big(v_{R}^{2} + v_{u}^{2}\Big)Y_{\nu}^{2} \\ 
m_{\phi_{d}\phi_{R}} &= -\frac{1}{\sqrt{2}} v_L Y_\nu \mu \\
m_{\phi_{u}\phi_R} &= v_L \Big(\frac{1}{\sqrt{2}} T_\nu  - v_{\eta} Y_x Y_\nu \Big)\\ 
m_{\phi_L\phi_R} &= \frac{1}{\sqrt{2}} v_u T_\nu  + Y_\nu \Big(- \frac{1}{\sqrt{2}} v_d \mu  - v_u v_{\eta} Y_x \Big)\\ 
m_{\phi_R\phi_R} &= {m_{\nu^c}^2}+\frac{1}{8} g_{BL} \Big(- \bar{g} \Big(- v_{u}^{2}  + v_{d}^{2} + v_{L}^{2}\Big) + g_{BL} \Big(2 v_{\bar{\eta}}^{2}  -2 v_{\eta}^{2}  - v_{L}^{2}  + v_{R}^{2}\Big)\Big)\nonumber \\ 
 &+\frac{1}{2} \Big(-2 \sqrt{2} v_{\eta} T_x  + 2 Y_x \Big(\Big(2 v_{\eta}^{2}  + v_{R}^{2}\Big)Y_x  + \sqrt{2} {\mu'} v_{\bar{\eta}} \Big) + \Big(v_{L}^{2} + v_{u}^{2}\Big)Y_{\nu}^{2} \Big)\\ 
m_{\phi_u\phi_{\eta}} &= v_L v_R Y_x Y_\nu \\
m_{\phi_L\phi_{\eta}} &=v_R v_u Y_x Y_\nu \\
m_{\phi_R\phi_{\eta}} &= \sqrt{2} v_R T_x  + v_L v_u Y_x Y_\nu \\ 
m_{\phi_R\phi_{\bar{\eta}}} &=\sqrt{2} {\mu'} v_R Y_x  \\
m_{\phi_{\eta}\phi_{\eta}} &= 2 v_{R}^{2} Y_{x}^{2}  + \frac{1}{4} g_{BL} \Big(\bar{g} \Big(v_{d}^{2}- v_{u}^{2}  + v_{L}^{2}\Big) + g_{BL} \Big(2 \Big( v_{\eta}^{2}- v_{\bar{\eta}}^{2} \Big) - v_{R}^{2}  + v_{L}^{2}\Big)\Big) + m_{\eta}^2 + {\mu'}^{2}\\ 
m_{\phi_{\bar{\eta}}\phi_{\bar{\eta}}} &= \frac{1}{4} g_{BL} \Big(- \bar{g} \Big(- v_{u}^{2}  + v_{d}^{2} + v_{L}^{2}\Big) + g_{BL} \Big(2 v_{\bar{\eta}}^{2}  -2 v_{\eta}^{2}  - v_{L}^{2}  + v_{R}^{2}\Big)\Big) + m_{\bar{\eta}}^2 + {\mu'}^{2}
\end{align} 

\item {\bf Charged slepton - charged Higgs} \\
In the basis $$ \left(H_d^-, H_u^{+,*}, \tilde{e}_L,
\tilde{e}_R\right),$$ 
the entries of the mass matrix read:
\begin{align} 
m_{H_d^-H_d^{-,*}} &= m_{H_d}^2+\frac{1}{8} \Big(\bar{g} g_{BL} \Big(2 \Big(- v_{\bar{\eta}}^{2}  + v_{\eta}^{2}\Big) - v_{R}^{2}  + v_{L}^{2}\Big) + \Big(g_{1}^{2} + \bar{g}^{2}\Big)\Big(- v_{u}^{2}  + v_{d}^{2} + v_{L}^{2}\Big) \Big. \nonumber \\ 
 & \Big.+ g_{2}^{2} \Big(- v_{L}^{2}  + v_{d}^{2} + v_{u}^{2}\Big)\Big)+\frac{1}{2} v_{L}^{2} Y_{e}^{2} +\mu^{2}\\ 
m_{H_d^{-b}H_u^+} &= \frac{1}{4} g_{2}^{2} v_d v_u  + B_{\mu} \\
m_{H_u^{+,*}H_u^+} &= m_{H_u}^2+\frac{1}{8} \Big(\Big(- \bar{g}^{2}  - g_{1}^{2} \Big)\Big(- v_{u}^{2}  + v_{d}^{2} + v_{L}^{2}\Big) + \bar{g} g_{BL} \Big(2 v_{\bar{\eta}}^{2}  -2 v_{\eta}^{2}  - v_{L}^{2}  + v_{R}^{2}\Big) \nonumber \\ 
 &+ g_{2}^{2} \Big(v_{d}^{2} + v_{L}^{2} + v_{u}^{2}\Big)\Big)+\frac{1}{2} v_{R}^{2} Y_{\nu}^{2} +\mu^{2}\\ 
m_{H_d^-\tilde{e}_L^*} &= -\frac{1}{2} v_d v_L Y_{e}^{2}  + \frac{1}{4} g_{2}^{2} v_d v_L  - \frac{1}{\sqrt{2}} v_R Y_\nu \mu \\ 
m_{H_u^{+,*}\tilde{e}_L^*} &= -\frac{1}{2} Y_\nu \Big(2 v_R v_{\eta} Y_x  + v_L v_u Y_\nu \Big) + \frac{1}{4} g_{2}^{2} v_L v_u  - \frac{1}{\sqrt{2}} v_R T_\nu \\ 
m_{\tilde{e}_L\tilde{e}_L^*} &= m_L^2+\frac{1}{8} \Big(\Big(g_{1}^{2} + \bar{g}^{2}\Big)\Big(v_{d}^{2} - v_{u}^{2} + v_{L}^{2}\Big)+g_{2}^{2} \Big(v_{L}^{2} - v_{d}^{2}  +  v_{u}^{2}\Big)+\frac{1}{2} \Big(v_{d}^{2} Y_{e}^{2}  + v_{R}^{2} Y_{\nu}^{2} \Big)\nonumber \\ 
 &+\bar{g} g_{BL} \Big(-2 v_{\bar{\eta}}^{2}  + 2 \Big(v_{L}^{2} + v_{\eta}^{2}\Big) - v_{R}^{2}  - v_{u}^{2}  + v_{d}^{2}\Big)+g_{BL}^{2} \Big(2 \Big(- v_{\bar{\eta}}^{2}  + v_{\eta}^{2}\Big) - v_{R}^{2}  + v_{L}^{2}\Big)\Big)\\ 
m_{H_d^-\tilde{e}_R^*} &= -\frac{1}{2} v_R v_u Y_e Y_\nu  - \frac{1}{\sqrt{2}} v_L T_e \\ 
m_{H_u^{+,*}\tilde{e}_R^*} &= -\frac{1}{2} Y_e \Big(\sqrt{2} v_L \mu  + v_d v_R Y_\nu \Big)\\ 
m_{\tilde{e}_L\tilde{e}_R^*} &= \frac{1}{\sqrt{2}} \Big(v_d T_e  - v_u Y_e \mu \Big)\\ 
m_{\tilde{e}_R\tilde{e}_R^*} &= m_E^2+\frac{1}{8} \Big(- \Big(2 \bar{g}  + g_{BL}\Big)\Big(\bar{g} \Big(- v_{u}^{2}  + v_{d}^{2} + v_{L}^{2}\Big) + g_{BL} \Big(-2 v_{\bar{\eta}}^{2}  + 2 v_{\eta}^{2}  - v_{R}^{2}  + v_{L}^{2}\Big)\Big)\nonumber \\ 
 & -2 g_{1}^{2} \Big(- v_{u}^{2}  + v_{d}^{2} + v_{L}^{2}\Big)\Big)+\frac{1}{2} \Big(v_{d}^{2} + v_{L}^{2}\Big)Y_{e}^{2} 
\end{align} 

We write only independent entries as the matrix is hermitian thus
$m^2_{ij}=(m^2_{ji})^*$.
\end{itemize}
\bibliography{BLSSM_vacuum.bib}
\bibliographystyle{h-physrev5}
\end{document}